\documentclass[manuscript,screen]{acmart}

\usepackage{color}
\usepackage{tcolorbox}

\usepackage{pifont, amsmath}

\usepackage{enumitem}
\usepackage[normalem]{ulem}
\usepackage{rotating}
\usepackage{color}
\usepackage{tabularray}
\usepackage{array}

\usepackage[normalem]{ulem}

\usepackage[resetlabels,labeled]{multibib}
\newcites{P}{Appendix A. Primary Studies}

\usepackage{graphicx, subcaption}
\usepackage{caption}
\AtBeginDocument{%
  }

\begin{document}

%%
%% The "title" command has an optional parameter,
%% allowing the author to define a "short title" to be used in page headers.
\title{Novice Developers' Perspectives on Adopting LLMs for Software Development: A Systematic Literature Review}

%%
%% The "author" command and its associated commands are used to define
%% the authors and their affiliations.
%% Of note is the shared affiliation of the first two authors, and the
%% "authornote" and "authornotemark" commands
%% used to denote shared contribution to the research.
\author{Samuel Ferino}
% \authornote{Both authors contributed equally to this research.}
\orcid{0000-0002-5484-1169}
\email{samuel.demouraferino@monash.edu}
\author{Rashina Hoda}
\orcid{0000-0001-5147-8096}
\email{rashina.hoda@monash.edu}
\author{John Grundy}
\orcid{0000-0003-4928-7076}
\email{john.grundy@monash.edu}
\affiliation{%
  \institution{Monash University}
  \city{Melbourne}
  \state{Victoria}
  \country{Australia}
}

\author{Christoph Treude}
\orcid{0000-0002-6919-2149}
\affiliation{%
  \institution{Singapore Management University}
  \city{Singapore}
  \country{Singapore}}
\email{ctreude@smu.edu.sg}

% \author{Valerie B\'eranger}
% \affiliation{%
%   \institution{Inria Paris-Rocquencourt}
%   \city{Rocquencourt}
%   \country{France}
% }

% \author{Aparna Patel}
% \affiliation{%
%  \institution{Rajiv Gandhi University}
%  \city{Doimukh}
%  \state{Arunachal Pradesh}
%  \country{India}}

% \author{Huifen Chan}
% \affiliation{%
%   \institution{Tsinghua University}
%   \city{Haidian Qu}
%   \state{Beijing Shi}
%   \country{China}}

% \author{Charles Palmer}
% \affiliation{%
%   \institution{Palmer Research Laboratories}
%   \city{San Antonio}
%   \state{Texas}
%   \country{USA}}
% \email{cpalmer@prl.com}

% \author{John Smith}
% \affiliation{%
%   \institution{The Th{\o}rv{\"a}ld Group}
%   \city{Hekla}
%   \country{Iceland}}
% \email{jsmith@affiliation.org}

% \author{Julius P. Kumquat}
% \affiliation{%
%   \institution{The Kumquat Consortium}
%   \city{New York}
%   \country{USA}}
% \email{jpkumquat@consortium.net}

%%
%% By default, the full list of authors will be used in the page
%% headers. Often, this list is too long, and will overlap
%% other information printed in the page headers. This command allows
%% the author to define a more concise list
%% of authors' names for this purpose.
\renewcommand{\shortauthors}{Ferino et al.}

%%
%% The abstract is a short summary of the work to be presented in the
%% article. have emerged in recent years.
\begin{abstract}
Following the rise of large language models (LLMs), many studies have emerged in recent years focusing on exploring the adoption of LLM-based tools for software development by novice developers: computer science/software engineering students and early-career industry developers with two years or less of professional experience. These studies have sought to understand the perspectives of novice developers on using these tools, a critical aspect of the successful adoption of LLMs in software engineering. To systematically collect and summarise these studies, we conducted a systematic literature review (SLR) following the guidelines by Kitchenham et al. on 80 primary studies published between April 2022 and June 2025 to answer four research questions (RQs). In answering RQ1, we categorised the study motivations and methodological approaches. In RQ2, we identified the software development tasks for which novice developers use LLMs. In RQ3, we categorised the advantages, challenges, and recommendations discussed in the studies. Finally, we discuss the study limitations and future research needs suggested in the primary studies in answering RQ4. Throughout the paper, we also indicate directions for future work and implications for software engineering researchers, educators, and developers. Our research artifacts are publicly available at \url{https://github.com/Samuellucas97/SupplementaryInfoPackage-SLR}. 
\end{abstract}

%%
%% The code below is generated by the tool at http://dl.acm.org/ccs.cfm.
%% Please copy and paste the code instead of the example below.
%%

%%
%% Keywords. The author(s) should pick words that accurately describe
%% the work being presented. Separate the keywords with commas.

% Large language models \sep Software engineering \sep  Junior software developers \sep  Human aspects \sep  Systematic literature review

% \received{20 February 2007}
% \received[revised]{12 March 2009}
% \received[accepted]{5 June 2009}

%%
%% This command processes the author and affiliation and title
%% information and builds the first part of the formatted document.
\maketitle

\section{Introduction}

\nociteP{*}

% \textcolor{red}{Samuel, this is a decent intro; but not an exciting one. The order of arguments is odd. Also, it tries to justify instead of motivate. Putting the definition first also feels defensive. Consider motivating why readers/reviewers should be concerned about novice devs perspectives and experiences of using LLM-based tools. Consider these key points, in this order: 1. widespread fear and speculations that LLMs will replace novice developers [cite refs from SD trends magazines, research papers; 2. Novice developers face an uncertain future [cite research/industry refs where novice devs themselves express these concerns] 3. Studies have been exploring this topic. Few SLRs exist [cite most highly relevant ones] BUT they do not do [add what our unique selling points are]. 4. Motivated by the above, we conducted an SLR that investigates x, y, z (key high level themes). 4a. Imporant to define novice. No universal definition exists, e.g., X says such and such, Y says such and such, add Begel quote, then explain OUR definition based on: what? a combined approach? 5. List key contributions. 6. Rest of the paper is structured as follows.}

According to Lenarduzzi et al., \textit{"Software engineering is one of the engineering fields with the highest inflow of junior engineers. The disproportion of junior and senior developers is increasing fast, and it puts a significant stress on the mentoring and tutoring process"}. This is motivated, for example, by the potential cost reduction compared to hiring experienced developers. However, the productivity of junior developers is not at the level of senior developers. LLM tools appear as a candidate alternative to enhance the productivity of novice developers \cite{cui:2024}, as well as to support senior developers in mentoring junior developers \cite{france:2024}. These tools trained with millions of code lines can support a variety of software development tasks such as code understanding, bug localisation, vulnerability detection, and test case generation \cite{hou:2024}, with minimal required human effort \cite{zhou:2024}. Studies are emerging showing the potential advantages related to LLM adoption \cite{ziegler:2024, cui:2024, ebert:2023}. Cui et al. \cite{cui:2024} conducted a controlled experiment with 4.867 software developers from Microsoft, Accenture, and an anonymous company. They found that the group of developers using GitHub Copilot achieved an increase of 26.08\% in the number of completed tasks weekly. Enterprise LLM adoption reports, such as McKinsey \cite{mayer:2025}, Kong \cite{kong:2025} and DORA \cite{dora:2024}, show an increase in companies adopting LLM tools \cite{mayer:2025}, where many companies are exploring the potential benefits (e.g., productivity, job satisfaction) for software development. On the other hand, there is concern that early career software developers might not have mastered the core competencies in software engineering (e.g., coding, and testing) to be able to use those tools properly \cite{kam:2025}.  Woodruff et al. \cite{woodruff:2024} also mention rising concerns among developers about future implications for their jobs.

In this context, Begel et al. \cite{begel:2008} raise an important point about university computer science (CS) / software engineering (SE) graduates becoming novice developers \textit{again} after joining the job market: 

\begin{quote}
\textit{"Software developers begin a transition from novice to expert at least twice in their careers – once in their first year of university computer science, and second when they start their first industrial job. Novice computer scientists in university learn to program, to design, and to test software. Novices in industry learn to edit, debug, and create code on a deadline while learning to communicate and interact appropriately with a large team of colleagues."} \cite{begel:2008}
\end{quote}

\noindent As soon as they graduate and join a company as junior software developers, they confront the bitter reality of the discrepancy between capstone projects and real-life projects \cite{craig:2018, begel:2008}. They also face the difference in expectations while joining the industry \cite{oguz:2019}. Alboaouh \cite{alboaouh:2018} argues that it may take two to three years for a graduate to adjust to industry standards and practices. This includes, for example, improving their technical skills (e.g., learning new frameworks and libraries) by working in large-scale software projects \cite{oguz:2019}. Thus, early career software developers are also still novice developers in these aspects.

Many studies in Software Engineering (SE) acknowledge the potential effects and implications specific to the population of novice developers and specifically target novice developers as study participants (e.g., \cite{gilson:2020, molnar:2024, ahmad:2024, craig:2018, mian:2022}). For instance, Zhao et al. \cite{zhao:2024} investigate the factors (e.g., newbie-manager pairing, job satisfaction) that influence early career software developers. They identified how crucial the early stage is as the foundation for a solid software developer's career, which may incur negative consequences for software developers and development teams if disregarded. In another study, Lenarduzzi et al. \cite{lenarduzzi:2020} identified that junior developers tend to \textit{"incur more code smells over other types of technical debt"}. Because of the profound shift the world has experienced due to the large language model (LLM)-driven revolution sparked by the release of ChatGPT in November 2022 \cite{teubner:2023, dwivedi:2023}, it is also necessary to investigate novice developers' perspectives. In the LLMs for Software Engineering (LLM4SE) research area, Tona et al. \cite{tona:2024} detected discrepancies between the perspectives of industry software engineers and SE students through an experimental study. They found that while industry software engineers see the risks LLM tools pose to \textit{"quality of work and professional practice"}, SE students basically perceive those tools as additional tools. That naive mindset regarding the risks of LLM adoption can lead to unforeseeable consequences, as evidenced by one of the participants in Mendes et al.'s study \citeP{P68:mendes2024}: \textit{"I've had a bad experience because I've uploaded code that I thought was good, but in production, it was unworkable, you know?"}. These situations may compromise the end-user experience (in the worst scenario) as well as contribute to the introduction of bugs.

Systematic literature reviews (SLRs) are emerging in the literature, exploring novice developers adopting  LLM tools \cite{cambaz:2024, pirzado:2024, raihan:2025, prather:2025}. However, they focus on understanding how LLM tools are used in computing education \cite{pirzado:2024, prather:2025, raihan:2025}, such as teaching and learning practices \cite{cambaz:2024} that novice developers utilised LLMs. These SLRs are related to our SLR while investigating novice developers' usage of LLMs, but our SLR differs in terms of purpose and population. These SLRs lack a comprehensive understanding of novice software developers' adoption (e.g., challenges, recommendations) and use of LLMs in SE activities, from the perspective of both CS/SE students and junior software developers. Table \ref{tab:relatedWork} summarises the differences between the related works and our SLR.

  \begin{table}
\centering
\scriptsize
\caption{State-of-the-art SLRs related to novice developers using LLMs.}
\label{tab:relatedWork}
\begin{tblr}{
  width = \linewidth,
  colspec = {Q[143]Q[32]Q[120]Q[128]Q[140]Q[80]Q[100]},
  row{even} = {c},
  row{1} = {c},
  row{3} = {c},
  row{5} = {c},
  row{6} = {c},
  row{8} = {c},
  cell{2}{3} = {r=4}{},
  cell{2}{4} = {r=4}{},
  cell{6}{1} = {r=2}{},
  cell{6}{2} = {r=2}{},
  cell{6}{5} = {r=2}{},
  cell{6}{6} = {r=2}{},
  cell{6}{7} = {r=2}{},
  row{7} = {c},
  hline{1-2,6,8} = {-}{},
}
\textbf{Reference}                                  & \textbf{Year} & \textbf{Scope}           & \textbf{Target population}         & \textbf{Coverage}     & \textbf{Time frame} & \textbf{\#Included papers} \\
Cambaz et al. \cite{cambaz:2024}   & 2024          & {LLMs in \\CS Education} & CS/SE Students~                    & SE Tasks, Perceptions & 2018-2023           & 21                         \\
Pirzado et al  \cite{pirzado:2024} & 2024          &                          &                                    & SE Tasks              & 2021-2024           & 72                         \\
Raihan et al.  \cite{raihan:2025}  & 2024          &                          &                                    & SE Tasks, Perceptions & 2019-2024           & 125                        \\
Prather et al. \cite{prather:2025} & 2025          &                          &                                    & SE Tasks, Perceptions & 2022-2024           & 71                         \\
Our work                                            & 2025          & Novice Developers        & CS/SE Students                     & SE Tasks, Perceptions & 2022-2025           & 80                         \\
                                                    &               &  \& LLM4SE  & \& Junior Developers &                       &                     &                            
\end{tblr}
\end{table}

Seeking to address this research gap, and also motivated by the rise of empirical studies exploring novice software developers' adoption and use of LLMs in SE activities, we conducted a systematic literature review aiming to comprehend novice software developers' perspectives on adopting LLM-based tools for software development tasks. Using the guidelines introduced by Kitchenham et al. \cite{kitchenham:2007, kitchenham:2022}, we selected 80 primary studies in our SLR from April 2022 to June 2025, identifying the study motivations and goals, methodological approaches, study limitations, study findings, and future research needs. During our analysis, we identified a variety of junior software developers' and CS/SE students' perceptions (e.g., usefulness, emotions, productivity), perceived benefits and challenges, general recommendations, and educational recommendations to support educators in the GenAI era. Although there is no universal understanding regarding junior software developers \cite{tona:2024}, we employ 2 years of professional experience as a threshold to identify them in the selected studies. Our analysis also covers the software development tasks in which novice software developers are adopting LLM-powered tools. Our work makes the following key contributions:

\begin{itemize}
    \item Analysis of publication trends covering domains, publication venues, research methods, and LLM-based tools; 
    \item Insights and guidance for IT professionals, educators, and researchers to help them improve their understanding of novice software developers' perception, potentially impacting  their decision about LLM adoption;
    \item A set of key research needs and recommendations for future research directions into Large Language Models for Software Engineering focusing on novice software developers.
\end{itemize}

This paper is organised in the following sections: Section \ref{sec:relatedWork} discusses the background knowledge and key related reviews. Section \ref{sec:methodology} presents our SLR research methodology.  Sections \ref{sec:rq1}, \ref{sec:rq2}, \ref{sec:rq3}, and \ref{sec:rq4} present the findings for each research question, respectively. Section \ref{sec:discussion} discusses key study results and suggests directions for future research. Section \ref{sec:threatsToValidity} discusses the study limitations, while section \ref{sec:conclusion} concludes this paper.

\section{Background and Related Work}\label{sec:relatedWork}

This section provides the background knowledge and discusses key secondary studies in LLM4SE, both in general and focused on novice developers, that are related to our work.

\subsection{Definition of Novice Software Developers}

As mentioned in the Introduction section, our understanding of novice developers includes both university students and industrial junior software developers. According to Tona et al., there is no universal understanding of the definition of junior developers \cite{tona:2024}. This is highlighted by Zhao et al. \cite{zhao:2024} when they employed five years as a threshold to define early career software developers. In the study by Niva et al. \cite{niva:2023}, they identified a job description for junior developers that mentioned the requirement of less than three years of professional experience. On the other hand, Gilson et al. \cite{gilson:2020} use third-year university students as a proxy for junior developers. This demonstrates how flexibly the concept of a junior developer is being addressed by the research community. Tona et al. \cite{tona:2024} use the following definition: \textit{"A Junior Software Engineer (Jr.) usually has 0-2 years of experience"}. Note that, since university students \textit{usually} have less than 2 years of professional experience, they would also fit this definition. University students may also have the opportunity to attend industry internships \cite{minnes:2021, kapoor:2020, kapoor:2019}. In our study, we employ 2 years of professional experience as a threshold to avoid ambiguity in this key term. We decided to set a threshold given that we observed during tests with our search string that most of the studies do not characterise participants' expertise, only providing years of experience. Ideally, studies should seek to characterise participants in terms of expertise, combining different mechanisms (e.g., self-assessed expertise) \cite{baltes:2018}. 

\subsection{Secondary Studies in LLM4SE}

Our analysis of the related reviews uncovered seven secondary studies focusing on LLM4SE. Hou et al. \cite{hou:2024} conducted an SLR focusing on understanding how LLMs can improve processes and outcomes, selecting 395 primary studies. Similar to our study, their findings revealed the different LLMs employed in software development tasks and strategies used to improve the performance of LLMs in SE. It also includes successful use cases for LLMs. Zheng et al. \cite{zheng:2025} also conducted an SLR in LLM4SE, selecting 123 primary studies. They discuss the research status of LLMs from the point of view of seven software engineering tasks: code generation, code summarisation, code translation, vulnerability detection, code evaluation, code management, Q\&A interaction, and other works. Their findings also present the performance and effectiveness of LLMs in many development tasks. Zhang et al. \cite{zhang:2024} conducted a systematic survey aiming to summarise the capabilities of LLMs and their effectiveness in SE. Their analysis of the 1,009 selected studies identified what software development tasks are facilitated by LLM tools and factors influencing LLM adoption in SE. Sasaki et al. \cite{sasaki:2024} conducted an SLR of prompt engineering patterns in SE, aiming to organise them into a taxonomy supporting LLM adoption in SE. Based on an analysis of 28 selected studies, their findings resulted in 21 prompt engineering patterns under five main categories. He et al. \cite{he:2025} conducted a systematic review of LLM-Based Multi-Agent Systems for SE, resulting in 71 relevant studies. Their findings include discussing the applications and capabilities of LLM-based multi-agent systems in development (e.g., software maintenance). Fan et al. \cite{fan:2023} provide a survey of the emerging areas in LLM4SE, revealing open research challenges regarding the adoption of LLM tools. However, they did not conduct a systematic literature review. We also found a tertiary study by Gormez et al. \cite{gormez:2024}. They conducted a systematic mapping study unveiling the capabilities and potential of LLMs in software development tasks. They analysed seven systematic literature reviews, identifying LLMs and the challenges faced while using them. Sergeyuk et al. \cite{sergeyuk:2025} conducted a systematic literature review on human-AI experience in Integrated Development Environment (IDE), identifying 89 primary studies. They also identified benefits and challenges related to the adoption of these tools. To summarise, none of these related secondary studies seek to understand novice software developers' context in using LLMs for Software Engineering.

\subsection{Secondary Studies of Novice Developers \& LLM4SE}

During our search in the literature for secondary studies on novice developers using LLMs for software development, we identified four related works. Cambaz et al. \cite{cambaz:2024} conducted a systematic literature review on the usage of LLM tools for code generation in teaching and learning programming, selecting twenty-one papers published from 2018 to 2023. They searched for terms related to AI tools, computing education, and software engineering in ACM DL, Scopus, and Google Scholar during their paper selection process.  They identified educational practices where novice developers utilise LLMs for code generation (e.g., automatic generation of students' assignments). Pirzado et al. \cite{pirzado:2024} conducted a systematic literature review to understand the extent of LLM adoption in computing education. They selected 72 studies from 2021 to 2024, focusing on LLMs used by CS students as coding and debugging assistants. They searched for terms related to AI tools (e.g., Codex), computing education (e.g., Computer Science students), code generation, code explanation, and challenges on IEEE Xplore, ACM DL, ScienceDirect, Web of Science, and SpringerLink during their paper selection process, which was reinforced by a snowballing process. They identified challenges regarding incorporating LLMs in computer education (e.g., lack of accuracy).  Raihan et al. \cite{raihan:2025} selected 125 primary studies from January 2019 to June 2024 during their systematic literature review on the impact of LLMs on computer education. They focused on capturing a general landscape, including the most commonly used programming languages and the LLMs being employed. During their paper selection, they searched for terms related to software engineering (e.g., software development), education (e.g., teaching), and LLMs on ACM DL, IEEE Xplore, Science Direct, Scopus, ACL Anthology\footnote{\url{https://aclanthology.org}}, Web of Science, and arXiv. They identified students' and instructors' sentiment regarding using LLMs in computer science education as mostly positive in the studies. Prather et al. \cite{prather:2025} selected 71 primary studies until May 2024 during their systematic literature review on how instructors integrate generative AI into computing classrooms. They identified a few software development tasks that CS students are using LLM tools (e.g., writing code). Their paper selection process involved searching for terms related to computing education (e.g., computer science education), LLM tools (e.g., ChatGPT), education (e.g., pedagogy), and research methods (e.g., interview) on ACM DL, IEEE Xplore, Scopus, ASEE Peer\footnote{\url{https://peer.asee.org}}, and arXiv. They identified CS students using LLMs for a few SE activities such as debugging, code generation, and code review.

In summary, previous secondary studies focused on understanding the adoption of LLM tools by novice developers in an educational context. None of these related secondary studies seeks to understand novice software developers' context in using LLMs for Software Engineering, combining both CS/SE students and early career industry developers. Although secondary research focused on novice developers using LLMs is growing, there is no SLR that summarises those novice developers' perceptions and the software development tasks they are using LLMs. Our SLR has a key focus on novice developers' perspectives, including analyses of empirical studies with CS/SE students and early career industry developers. Table \ref{tab:relatedWork} summarises the differences between the related works and our SLR.

\section{Methodology} \label{sec:methodology}

\begin{figure}[ht]
    \centering
    \includegraphics[width=0.9\linewidth]{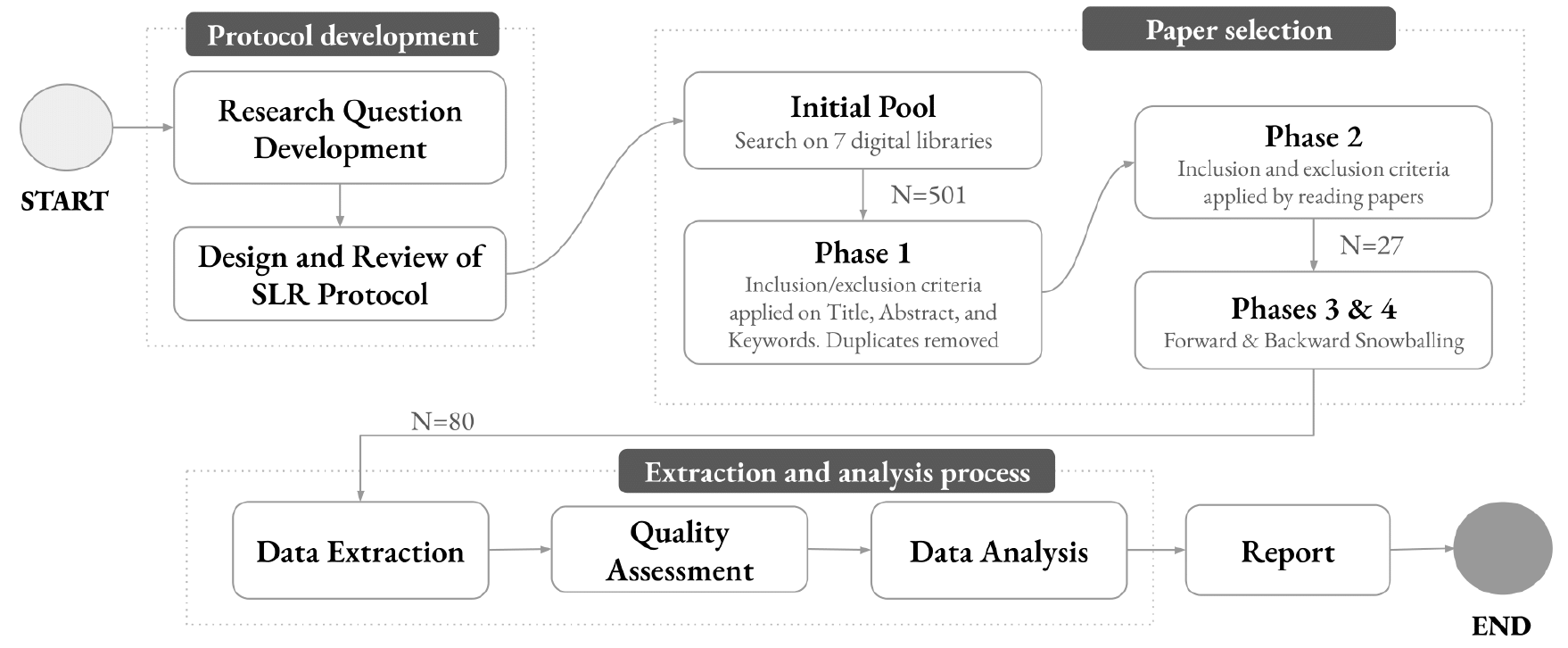}
    \caption{Systematic Literature Review process applied based on Kitchenham \cite{kitchenham:2007, kitchenham:2022}.}
    \label{fig:SLRStudyMethodology}
\end{figure}

Figure \ref{fig:SLRStudyMethodology} shows an overview of the SLR research methodology. Initially, the first author developed a preliminary SLR protocol, which was polished through many discussions with PhD supervisors and other SLR experts. The first author used Parsifal\footnote{\url{https://parsif.al}} - a platform focused on supporting SLR studies -  to support the SLR study design and execution. We also used spreadsheets for the snowballing process, since Parsifal does not support snowballing, and for data extraction and analysis. The first author applied the search string to the seven databases, filtered the relevant studies, and extracted and analysed the data by synthesising tables and figures. Each step conducted by the first author was done under the guidance of his PhD supervisors. We detail the research methodology in the following subsections. The research artifacts (data synthesis spreadsheet, data extraction form, SLR Protocol) are available here\footnote{\url{https://github.com/Samuellucas97/SupplementaryInfoPackage-SLR}}.  

\subsection{Research Questions}

To develop our research questions (RQs), we employed the PICOC framework (population, intervention, comparison, outcomes, and context), described in Table \ref{tab:picoc}. Wohlin et al. \cite{wohlin:2012} suggested the adoption of the PICOC framework as a foundation to develop well-defined research questions for SLRs. Other research questions were added as a result of discussions between the first author and his PhD supervisors. Thus, our study proposes to address the following research questions:

\begin{table}[ht]
\centering

\scriptsize
\caption{PICOC for Research Questions}
\begin{tblr}{
  width = \linewidth,
  column{1} = {c},
  cell{2}{1} = {c},
  cell{3}{1} = {c},
  cell{4}{1} = {c},
  cell{5}{1} = {c},
  hline{1,6} = {-}{0.08em},
}
\textbf{Population}   & Novice software developers                                                                                   \\
\textbf{Intervention} & Large Language Models (LLMs)                                                                                 \\
\textbf{Comparison}   & N/A                                                                                                          \\
\textbf{Outcome}      & {Novice software developers’ perceptions, challenges and recommendations regarding effects\\ on LLM adoption for software development} \\
\textbf{Context}      & Software Engineering                                                                                         
\end{tblr}
\label{tab:picoc}
\end{table}

\begin{itemize}
    \item \textbf{RQ1}. \textit{What are the \textbf{motivations} and \textbf{methodological approaches} behind each primary study to explore how novice software developers adopt LLM-based tools for software development tasks?} -- This RQ examines the primary goals, objectives, motivations, and methodologies employed by researchers to identify human aspects of novice software developers who adopt LLM-based tools for software development tasks. For example, we examine whether the study was conducted in an academic or industry setting. We also examine their strategies to categorise less experienced developers, i.e. novice software developers.

    \item \textbf{RQ2}. \textit{\textbf{What key software development tasks novice} developers are using LLM-based tools for?} -- In this RQ, we examine the software development tasks (e.g., coding, debugging, tests) that novice developers have been supported in by LLM-based tools. We also identify which LLM-based tools are being used by the novice software developers/team.
 
    \item \textbf{RQ3}. \textit{What are the \textbf{perceptions} and \textbf{experiences} of novice software developers when using LLM-based tools?} 

        \begin{itemize}
            \item[] \textbf{RQ3a.} \textit{What are the perceived and experienced \textbf{advantages/opportunities} of novice software developers when using LLM-based tools?} 
            \item[] \textbf{RQ3b.} \textit{What are the perceived and experienced \textbf{challenges/limitations} faced by novice software developers while using LLM-based tools?} 
            \item[] \textbf{RQ3c.} \textit{What are the \textbf{recommendations/best practices} suggested by novice software developers while using LLM-based tools?}
        \end{itemize}
        
\noindent In this RQ, we examine the novice software developers' perceptions involving benefits, challenges, limitations, and recommendations based on their experience using those LLM-based tools. 

    \item \textbf{RQ4}. \textit{What are the \textbf{limitations} and \textbf{recommendations for future research} that we can distil based on the primary studies?} -- In this RQ, we analyse the studies in terms of key contributions, limitations (e.g., in evaluation, participants), and future research recommended by the authors. Based on this, we suggest future work areas.

\end{itemize}

\subsection{Search Strategy}

\subsubsection{Search String}

We built our base search string from the PICOC framework (See Table \ref{tab:picoc}). We included relevant LLM-based tools such as ChatGPT and Copilot, and terms related to software practitioner roles. Based on the common and relevant database sources used in software engineering literature reviews (e.g., \cite{dybaa:2008, salleh:2011}), we decided to include the following six well-known database sources: ACM Digital Library, IEEE Xplore, SpringerLink, Scopus, ScienceDirect, and Wiley. We also included arXiv because research in LLM4SE is an emerging topic, and potentially relevant studies would be under review. Our search was limited to papers published in 2022, the year that ChatGPT became available for the general public. We refined our base search string following the requirements and setups described by each database. The refinement process was performed several times by applying the search string to the data sources and reading the titles, abstracts, and keywords of some papers. It was necessary to create smaller combinations of our search string for the ScienceDirect database due to word limitations. The research artifacts contain the final search strings for each database. Below, we present the base string:

\begin{itemize}
    \item \textit{("LLM" OR "large language model*" OR "ChatGPT" OR "Copilot" OR "Generative AI" OR "Conversational AI" OR "Chatbot*") AND (("junior*" OR "novice*") AND ("software developer*" OR "software engineer*" OR "software practitioner*" OR "programmer*" OR "developer*"))}
\end{itemize}

\subsubsection{Inclusion and Exclusion Criteria}

During the preparation of this SLR, we defined the inclusion and exclusion criteria. Table \ref{tab:inclusioExclusionCriteria} shows the inclusion and exclusion criteria adopted during paper selection.

\begin{table}[ht]
\centering
\scriptsize
\caption{Inclusion and Exclusion criteria}
\label{tab:inclusioExclusionCriteria}
\begin{tblr}{
  width = \linewidth,
  hline{1-2,10-11,16} = {-}{},
}
\textbf{ID} & \textbf{Inclusion criteria}                                                                                                                       \\
I01         & The paper is about  novice developers using LLM-based tools, including junior developers (0-2 years of industry experience) and CS/SE students \\
I02         & The paper must answer at least one of the RQs                                                                                                     \\
I03         & The paper is written in English                                                                                                                   \\
I04         & The full-text is accessible                                                                                                                       \\
I05         & The paper is not a duplication of others                                                                                                          \\
I06         & The paper was published in journals, conferences, and workshops                                                                                   \\
I07         & The paper is an empirical study                                                                                                                   \\
I08         & The paper was published from 2022, when ChatGPT became available for public access                                                                \\
\textbf{ID} & \textbf{Exclusion criteria}                                                                                                                       \\
E01         & Short papers that are less than four pages                                                                                                        \\
E02         & Papers based only on authors’ personal views without supporting data                                                                              \\
E03         & Conference or workshop papers if an extended journal version of the same paper exists                                                             \\
E04         & Non-primary studies (Secondary or Tertiary Studies)                                                                                               \\
E05         & Papers about educational contexts not including Computer Science and SE students' perceptions about using LLM-based tools                         
\end{tblr}
\end{table}

\subsection{Paper Selection}

The selection was structured using the following steps:

\begin{itemize}
    \item \textbf{Initial Pool:} The search strings were executed across the seven data sources in August (2024), retrieving 501 BibTex references. The BibTeX references containing title, abstract, and keywords were uploaded to the Parsifal platform.  
    \item \textbf{Phase 1:} Then, the papers were filtered by their title, abstract, and keywords, while applying the inclusion and exclusion criteria. We removed 40 duplicated papers using Parsifal's duplicate papers detection feature. We also decided to keep the papers that we found difficult to decide based only on their titles, abstracts, and keywords. At the end of this phase, 53 papers were chosen.
    \item \textbf{Phase 2:} The papers were filtered by reading in full while applying the inclusion \& exclusion criteria. It resulted in 25 primary studies. Table \ref{tab:breakDown} shows details regarding paper count according to each data source. This phase was concluded in September (2024).
    
    \item \textbf{Phase 3 (Snowballing - Round 1):} Manual search was employed using both \textit{backward} and \textit{forward snowballing} techniques over the 25 primary studies in October (2024). It was utilised Google Scholar during this phase. According to Wohlin \cite{wohlin:2014}, snowballing in SLRs is a key technique to complement the automated search on databases, reducing the risk of relevant studies not being included in the paper selection.
    The snowballing process was organised into four sub-phases (3.1, 3.2, 3.3, and 3.4). For sub-phase 3.1, the papers were filtered by title. For sub-phase 3.2, the papers were filtered based on the abstract and keywords. For sub-phase 3.3, the papers were filtered by skimming the introduction, methodology, results, and conclusion sections. In sub-phase 3.4, we read in full. At the end of this phase, 31 papers were selected, adding to the collection of primary studies. Table \ref{tab:breakDown} presents the paper count for each sub-phase of the snowballing process.
    
    \item \textbf{Phase 4 (Snowballing - Round 2):} Given the fast-paced research area, we conducted a second round of forwarding snowballing in June (2025) over the 56 primary studies selected in previous phases. We conducted phase 4 similarly to phase 3 (e.g., skimming title, abstract and keywords). It resulted in an additional 33 papers. 

    \item \textbf{Phase 5:} We revised the 89 selected papers to ensure that all papers, including industry junior software developers, follow the definition - professionals between 0-2 years of experience. That was necessary because we started the paper selection process using 0-5 years of professional experience as the threshold to categorise junior developers. During this process, we removed nine papers that we previously categorised as junior developers category under 0-5 years of experience. Thus, the paper selection process results in 80 primary studies. 
\end{itemize}

\begin{table}[ht]
\centering
\scriptsize
\caption{Breakdown of the paper count.}
\label{tab:breakDown}
\begin{tblr}{
  % width = \linewidth,
  % colspec = {Q[56]Q[137]Q[113]Q[42]Q[42]Q[56]Q[206]Q[96]Q[46]Q[46]Q[42]Q[42]},
  cells = {c},
  cell{1}{4} = {c=2}{0.084\linewidth},
  cell{1}{9} = {c=4}{0.176\linewidth},
  cell{3}{1} = {r=8}{},
  cell{3}{6} = {r=8}{},
  cell{3}{7} = {r=2}{},
  cell{3}{8} = {r=2}{},
  cell{3}{9} = {r=2}{},
  cell{3}{10} = {r=2}{},
  cell{3}{11} = {r=2}{},
  cell{3}{12} = {r=2}{},
  cell{7}{9} = {c=4}{0.176\linewidth},
  cell{9}{7} = {r=2}{},
  cell{9}{8} = {r=2}{},
  cell{9}{9} = {r=2}{},
  cell{9}{10} = {r=2}{},
  cell{9}{11} = {r=2}{},
  cell{9}{12} = {r=2}{},
  cell{11}{1} = {c=7}{0.651\linewidth},
  cell{11}{8} = {c=5}{0.272\linewidth},
  vline{6} = {3-10}{},
  hline{1,12} = {-}{0.08em},
  hline{3} = {1-5,7-12}{},
  hline{7} = {8-12}{},
  hline{9} = {7-12}{},
  hline{11} = {-}{},
}
                                                                                &                   &                       & \textbf{Phase} &             &                                                &                            &                       & \textbf{Phase} &              &              &              \\
                                                                                & \textbf{Resource} & \textbf{Initial Pool} & \textbf{1}     & \textbf{2}  &                                                & \textbf{Resource}          & \textbf{Initial Pool} & \textbf{3.1}   & \textbf{3.2} & \textbf{3.3} & \textbf{3.4} \\
\begin{sideways}\textsc{Primary Search}\end{sideways}                                    & ACM DL            & 184                   & 23             & 13          & \begin{sideways}\textsc{Secondary Search}\end{sideways} & Backward Snowballing       & 1645                  & 67             & 18           & 12           & 11           \\
                                                                                & IEEE Xplore       & 9                     & 3              & 2           &                                                &                            &                       &                &              &              &              \\
                                                                                & Springer          & 70                    & 5              & 2           &                                                & Forward Snowballing        & 924                   & 75             & 41           & 27           & 20           \\
                                                                                & Wiley             & 27                    & 1              & 0           &                                                & \textbf{COUNT}             & 2569                  & 142            & 59           & 39           & \textbf{31}  \\
                                                                                & Scopus            & 51                    & 8              & 5           &                                                &                            &                       & \textbf{Phase} &              &              &              \\
                                                                                & ScienceDirect     & 107                   & 7              & 1           &                                                & \textbf{\textbf{Resource}} & \textbf{Initial Pool} & \textbf{4.1}   & \textbf{4.2} & \textbf{4.3} & \textbf{4.4} \\
                                                                                & arXiv             & 53                    & 6              & 2           &                                                & Forward Snowballing        & 3963                  & 136            & 106          & 68           & \textbf{33 } \\
                                                                                & \textbf{COUNT}    & 501                   & 53             & \textbf{25} &                                                &                            &                       &                &              &              &              \\
\textbf{TOTAL FINAL PAPER COUNT} (Primary Search + Secondary Search - Phase 5)~ &                   &                       &                &             &                                                &                            & \textbf{\textbf{80}}  &                &              &              &              
\end{tblr}
\end{table}

\begin{figure}[ht]
    \centering
    \begin{subfigure}{0.45\textwidth}
        \centering
        \includegraphics[width=\textwidth]{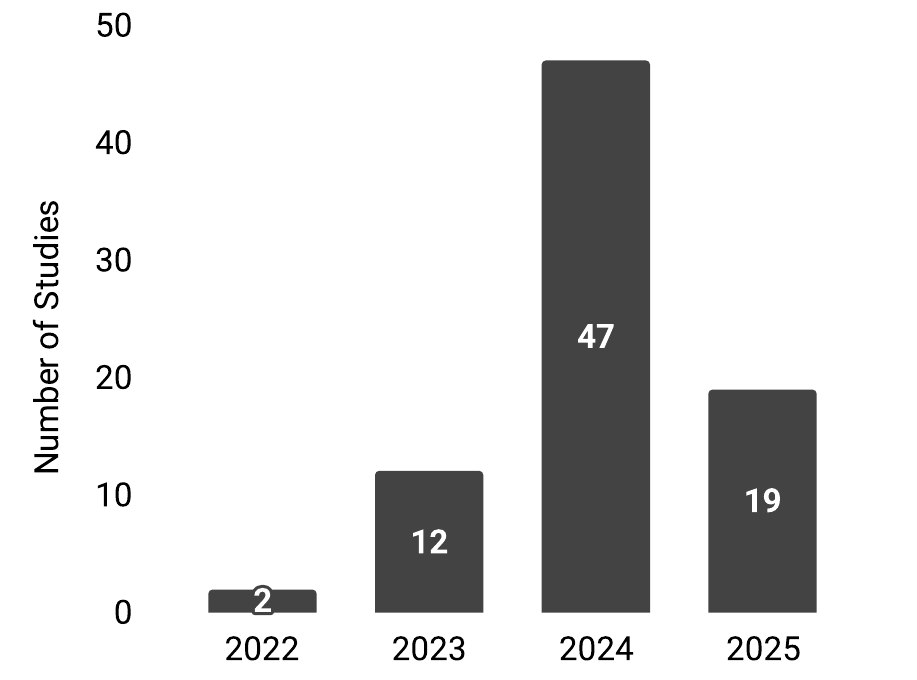}
        \caption{Distribution of the studies over years.}
        \label{fig:publicationsPerType}
    \end{subfigure}
    \hfill
    \begin{subfigure}{0.45\textwidth}
        \centering
        \includegraphics[width=\textwidth]{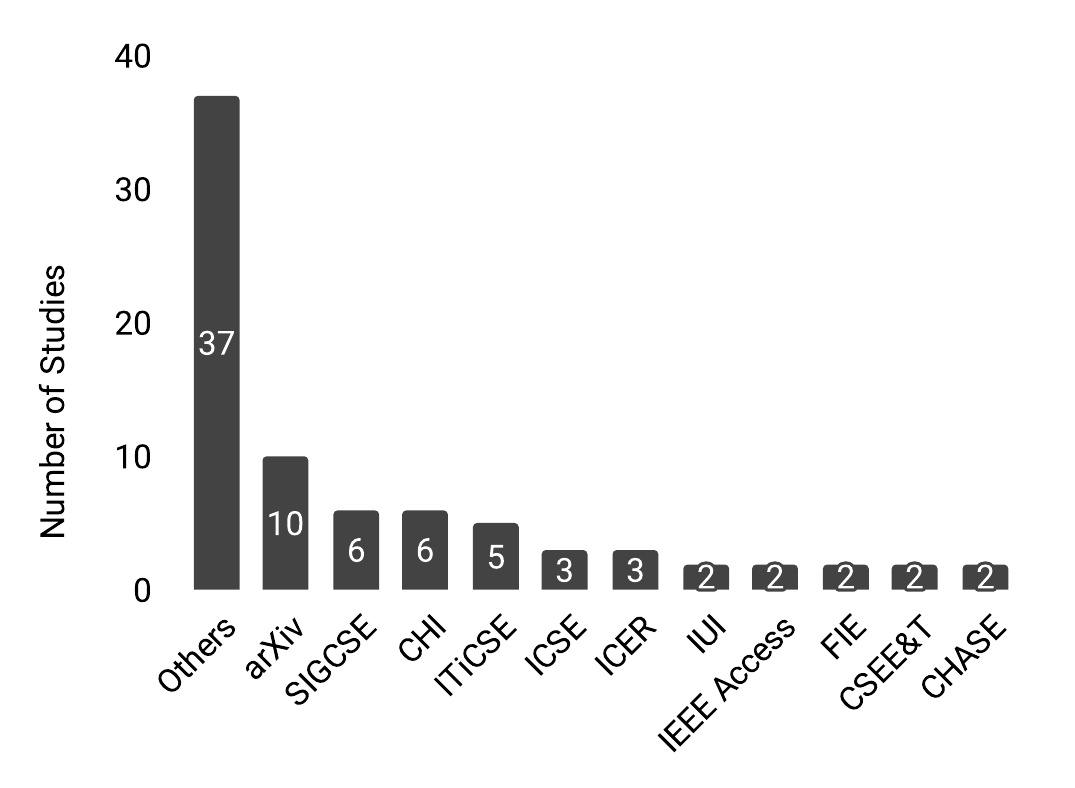}
        \caption{Distribution of the studies across venues.}
        \label{fig:publicationVenues}
    \end{subfigure}
    \caption{Overview of the 80 primary studies.}
    \label{fig:sidebyside}
\end{figure}

We obtained 80 primary studies from our paper selection process. The ACM digital library was the database source from which we retrieved the highest number of relevant papers - 16.2\%  (n = 13). At the same time, the snowballing forwarding approach exceeds the ACM DL by returning the highest number of relevant papers - 66.2\% (n = 53). The majority of the primary studies are conference papers (67.1\%, n=47). Our selected papers also include ten papers published on arXiv. Figure \ref{fig:publicationsPerType} shows the selected papers, including papers from 2022 to 2025. From 2022 to 2024, there is a consistent trend in the number of publications by year. It is already past the halfway point of 2025 (the current year), and the number of publications in 2025 has already surpassed those in 2023.

Figure \ref{fig:publicationVenues} shows the distribution of the publication venues, highlighting the venues with more than one publication. In terms of venues, 58.7\% of the studies were published in a CORE-ranked conference (A*, A, B, Australasian B, C, or National: Romania) or a ranked journal (Q1 or Q2). The Conference on Human Factors in Computing Systems (CHI) - the most prestigious conference in the field of Human-Computer Interaction - and the Technical Symposium on Computer Science Education (SIGCSE). For journal publications, we identify a few publications in HCI-focused journals such as ACM Transactions on Computer-Human Interaction (TOCHI) and Proceedings of the ACM on Human-Computer Interaction, but there is also a publication in the ACM Transactions on Software Engineering and Methodology (TOSEM) and another in the Information and Software Technology (IST), general-purpose software engineering journals.

\begin{figure}[ht]
    \centering
    \includegraphics[width=0.75\linewidth]{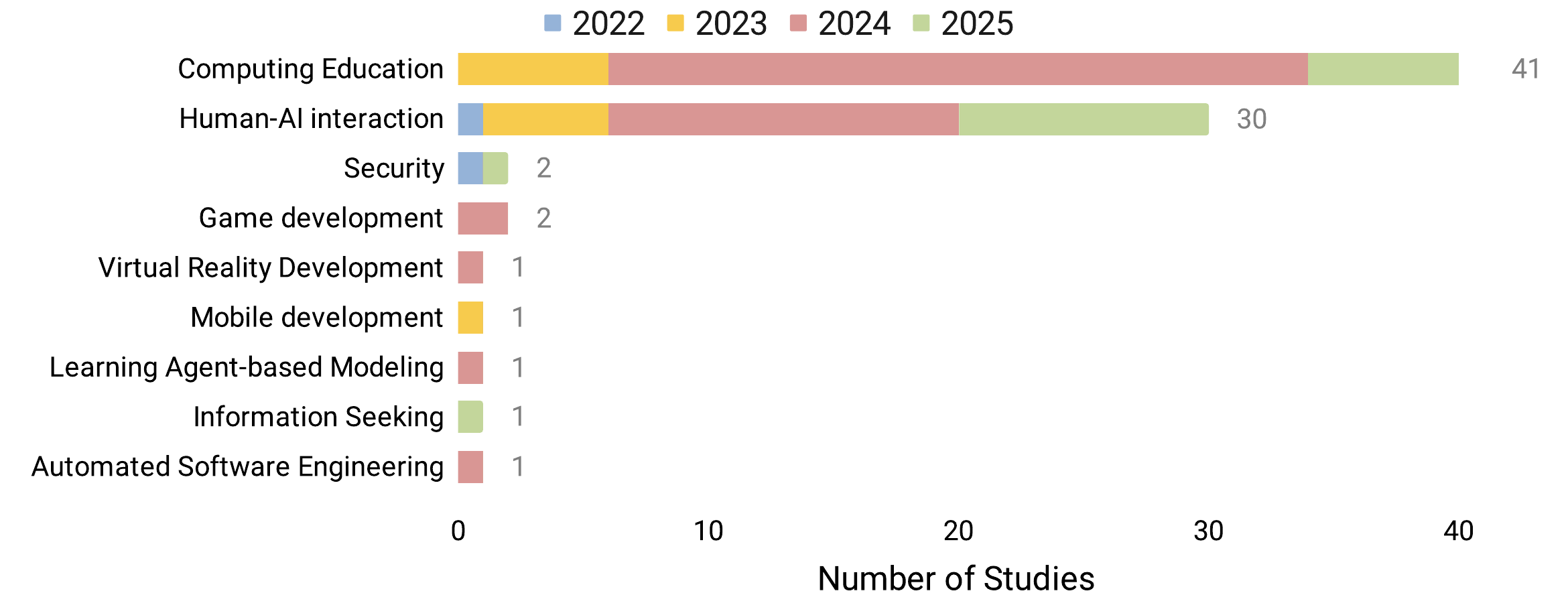}
    \caption{Distribution of domains over years.}
    \label{fig:publicationDomainsOverTime}
\end{figure}

Figure \ref{fig:publicationDomainsOverTime} presents the distribution of publication domains over the years, based on analysis of paper title, abstract, and keywords. Computing Education and Human-AI Interaction are the two most relevant domains, encompassing 88.7\% (n = 71) of the selected primary studies. We perceive a trend in publications focused on those two domains. However, a diversity of domains is emerging in 2024 (e.g., Game Development, Virtual Reality Development).

\subsection{Data Extraction and Analysis}

 The information necessary to address the RQs was extracted by inserting the data from each paper into a Google form composed of 40 questions, organised in the following 5 sections: 

\begin{enumerate}
    \item general information (e.g., paper title, paper's authors, year, venue);
    \item motivations and methodological approaches (e.g., study goal, research question);
    \item key software development tasks (e.g., key software development tasks, LLM-based tools being used);
    \item perceptions about LLM4SE (e.g., benefits, challenges);
    \item limitations and future research needs (e.g., main findings and limitations).
\end{enumerate}

This final version of the data extraction form was the result of adjustments after a pilot test with 21 papers. Our definition of novice developers encompasses both university students and industry junior software developers (0-2 years). The novice developers' perceptions in the studies with other study participant roles (e.g., senior developers, instructors) were collected by cross-checking with participant demographics described in the paper or supplementary material. For instance, Russo's study \citeP{P8:russo2024} comprehends novice and experienced developers as study participants; however, we only extracted the novice developers' perspectives.

The first author conducted a descriptive statistical analysis of the data using graphs and tables under the supervision of the other authors. Categories for the study motivations were created by first summarising the data extracted from the papers, allowing familiarisation with similarities in the study goals and motivations. Then they were grouped by categories and sub-categories. We did something similar during the categorisation of the benefits, challenges, recommendations, study limitations, and study recommendations for future work.

\subsection{Paper Assessment} 

Seeking to facilitate the understanding of the structure of our selected studies, we performed a paper assessment based on a yes-no-partial evaluation system. Our paper assessment strategy follows the literature on software engineering \cite{hoda:2017, kitchenham:2010}. Table \ref{tab:qualityAssessmentCriteria} shows our paper quality assessment strategy based on eight predefined questions. Similarly to Khalajzadeh and Grundy \cite{khalajzadeh:2024}, we employed the Computing Research and Education Association of Australasia (CORE\footnote{\url{https://portal.core.edu.au/conf-ranks}}) Conference and the Scimago Journal Rankings\footnote{\url{https://www.scimagojr.com}} to identify reputable venues (QA8). The results of the assessment are available in the supplementary package. Note that arXiv papers (not published) will receive \textit{No} in QA8. Papers published in conferences not ranked in CORE but have clear sponsorship by ACM or IEEE, traditional computing organisations, will receive \textit{Partially}, such as IEEE/ACM International Conference on Cooperative and Human Aspects of Software Engineering (CHASE). We decided to follow some SLRs (e.g., \cite{kitchenham:2009, salleh:2011, hidellaarachchi:2021, naveed:2024}) in not excluding papers based on quality assessment.

\begin{table}
\centering
\scriptsize
\caption{Paper quality checklist criteria.}
\begin{tblr}{
  width = \linewidth,
  colspec = {},
  row{1} = {c},
  hline{1,10} = {-}{0.08em},
  hline{2} = {-}{},
}
\textbf{ID} & \textbf{Criterion}                                                                              \\
QA1         & Is the paper highly relevant to the proposed SLR?                                               \\
QA2         & Is there a clear statement of the aim of the research?                                          \\
QA3         & Is there a review of key past work?                                                             \\
QA4         & Is there a clear research methodology which aligns with the key research questions of the study?    \\
QA5         & Does the paper provide sufficient information on data collection and data analysis of the research? \\
QA6         & Are the findings of the research clearly stated and supported by the research questions?        \\
QA7         & Does the paper provide limitations, summary and future work of the research?                    \\
QA8         & Is the paper published in a reputable venue?                                                    
\end{tblr}
\label{tab:qualityAssessmentCriteria}
\end{table}

\section{RQ1: What are the motivations and methodological approaches behind each primary study to explore how novice software developers adopt LLM-based tools for software development tasks?} \label{sec:rq1}

This section contains the findings regarding the first research question. We first present the study motivations, goals, and objectives identified in the primary studies, followed by the study methodologies and data analysis techniques.

\subsection{Study motivations, goals, and objectives}

Through our analysis of the motivations, goals, and objectives presented in each of the 80 primary studies, we classified them into four categories: integrating LLMs in SE and its implications, integrating LLMs in SE Education, with its implications, and integrating LLMs in specific industry domains. We identified 29 primary studies that explored novice developers' perceptions and attitudes (e.g., trust) towards LLMs, as well as their effectiveness in improving the productivity of software engineering tasks. For instance, Yang et al. \citeP{P36:yang2024} sought to understand the impact of LLM tools during debugging. We found 51 primary studies focusing on the integration of LLMs into Computer Education, including aspects such as perceptions, attitudes, advantages, and challenges. For instance, Shah et al. \citeP{P50:shah2025} seek to understand how CS students use GitHub Copilot. We found two primary studies investigating the adoption in specific industries. For instance, Boucher et al. \citeP{P14:boucher2024} seek to understand the impact of LLM adoption on game development. In this sense, we suggest that future research explore other domains, especially those that impose restrictions on privacy, such as government and finance. \label{rc:restrictDomains}

\subsection{Study methodologies and data analysis techniques}
\begin{figure}[ht]
    \centering
    \begin{subfigure}{0.45\textwidth}
        \centering
        \includegraphics[width=\textwidth]{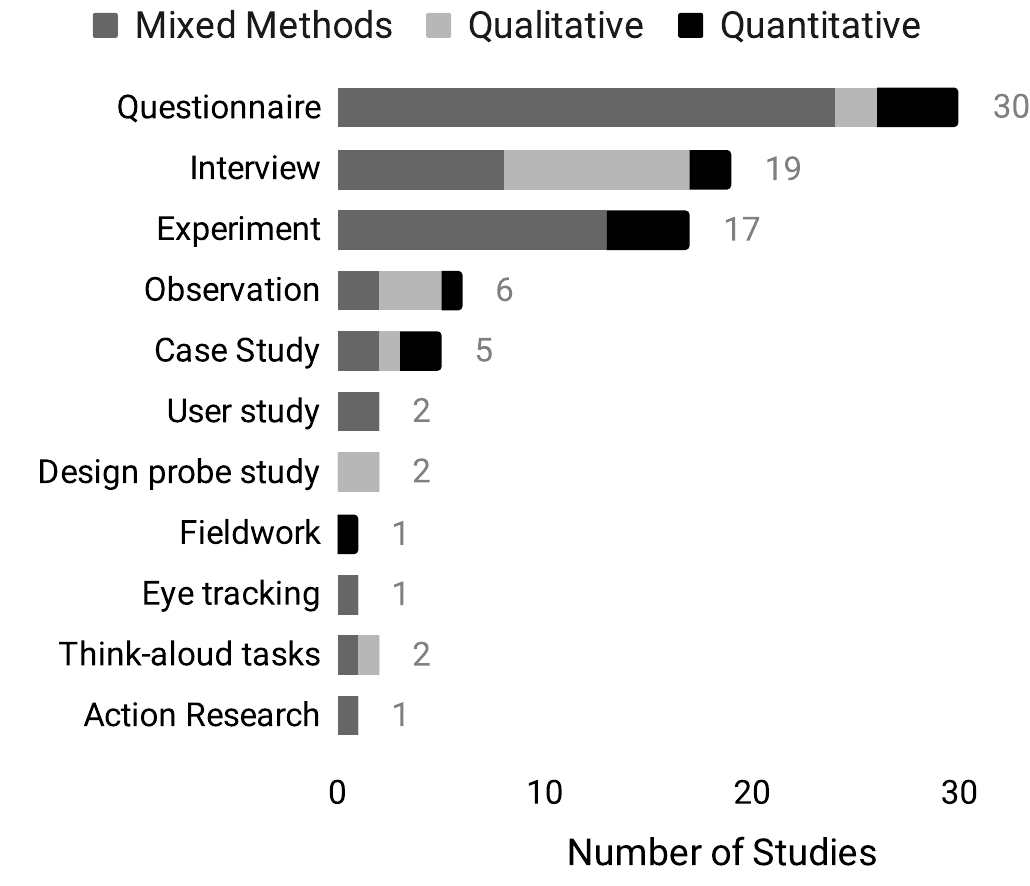}
        \caption{Distribution of research methods.}
        \label{fig:researchMethodOverType}
    \end{subfigure}
    \hfill
    \begin{subfigure}{0.45\textwidth}
        \centering
        \includegraphics[width=\textwidth]{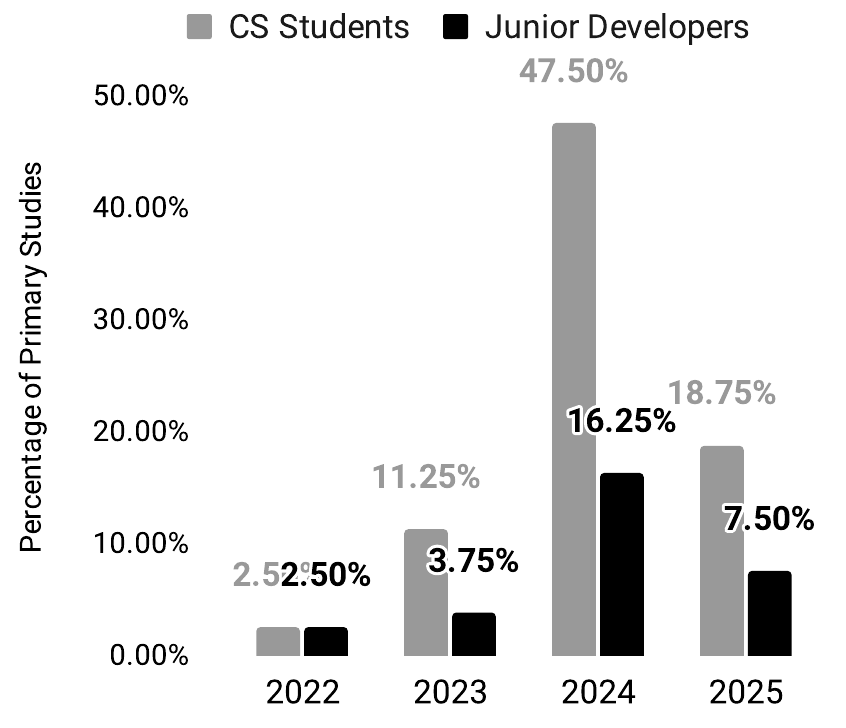}
        \caption{Distribution of the study participants.}
        \label{fig:studyParticipants}
    \end{subfigure}
    \caption{Overview of methodologies across the 80 primary studies.}
\end{figure}

Figure \ref{fig:researchMethodOverType} shows the distribution of the research methods in the 80 primary studies. Questionnaires (30 studies, 37.5\%) and interviews (19 studies, 23.7\%) were the most recurrent data collection methods. The interviews commonly followed the semi-structured format, but there is also Perry et al.'s study \citeP{P20:perry2023} and Choudhuri et al. \citeP{P62:choudhuri2025} where they employed retrospective interviews and reflective interviews, respectively. Most of the selected studies (49, 61.3\%) used mixed data analysis methods, while 18 studies (22.5\%) used qualitative methods, and 13 studies (16.3\%) used quantitative methods. While analysing the Figure \ref{fig:researchMethodOverType}, we noticed the potential of researchers to explore the implications of novice developers adopting LLM tools using the less commonly used research methods. For instance, researchers could replicate the think aloud-based study by Salerno et al. \cite{salerno:2024}, which explored challenges faced by novice developers when installing tools, in the context of LLM tools. Researchers could also replicate the study by Silva et al. \cite{silva:2023}, which investigated novice developers' code comprehension using eye tracking. Thematic analysis (36 studies, 48.2\%) and grounded theory (5 studies, 6.2\%) are the most commonly used qualitative data analysis techniques. Regarding quantitative data analysis techniques, we identified a variety of 33 techniques (e.g., student t-test, Mann-Whitney U test). Descriptive analysis was the most commonly employed quantitative data analysis technique - 33 studies, 41.2\%. Figure \ref{fig:studyParticipants} shows the distribution of study participants: CS/SE students and industrial junior software developers. Most of the studies investigate novice developers by using university students. The literature lacks studies in industry settings that are more realistic. \label{rc:industrySettings}

\begin{tcolorbox}[float=htpb!,colback=gray!5!white,colframe=gray!75!black,title=RQ1.  What are the motivations and methodological approaches behind each primary study to explore how novice software developers adopt LLM-based tools for software development tasks?]
We classified the goal stated in the primary studies into four main categories: integrating LLMs in SE (36.2\%), integrating LLMs in CS Education (63.7\%), and integrating LLM tools in specific industry domains (2\%). Questionnaires, interviews, and experiments were the most common data collection techniques, and both qualitative and quantitative methods were used for data analysis. Most studies have been done with academia-based projects. 
\end{tcolorbox}

\section{RQ2: What key software development tasks are novice developers using LLM-based tools for?}  \label{sec:rq2}

This section contains the findings regarding the second research question. While we focus on novice developers, we provide an analysis of the use of LLMs in different SE activities similar to Hou et al. \cite{hou:2024}, according to the SE activities in the Software Development Life Cycle (SDLC). The SDLC is organised in the following SE activities: requirement engineering, software design, software development, software quality assurance, software maintenance, and software management. They can be sequential or iterative depending on the SE methodology; however, the activities are more or less the same. Table \ref{tab:seActitivies} shows the distribution of SE tasks across 75 selected papers, including the number of paper occurrences. We did not identify reports of novice developers using LLMs for software project management tasks. According to the literature in SE, project management includes tasks such as effort estimation \cite{cabrero:2024, kula:2021, molokken:2003, jorgensen:2001} and task prioritisation \cite{xuan:2012, hujainah:2018}. Kula et al. \cite{kula:2021} and Molokken et al. highlighted that task effort estimation is heavily based on experts' past experiences. Since novice developers lack prior experience, they can be overly optimistic when estimating task effort \cite{jorgensen:2020}. A similar situation happens in task prioritisation \cite{hujainah:2018}. Based on this, there is a research gap involving investigations on the potential of LLM tools to support novice developers in effort estimation and task prioritisation. \label{rc:effortEstimationAndTaskPrioritisation}

\begin{table}
\centering
\caption{SE Activities in which LLM tools are employed by novice developers.}
\label{tab:seActitivies}
\begin{tblr}{
  width = \linewidth,
  colspec = {Q[213]Q[375]Q[342]},
  row{1} = {r},
  cell{1}{2} = {c=2}{0.717\linewidth,c},
  cell{2}{1} = {r=6}{r},
  cell{8}{1} = {r=9}{r},
  cell{17}{1} = {r=3}{r},
  vline{1-2, 4} = {1-19}{},
  hline{1,20} = {-}{0.08em},
  hline{2,8,17} = {-}{},
}
\textbf{SE Activity}                                         & \textbf{SE Tasks }                              &                                               \\
{Requirement Engineering \& \\Software Design     }            & brainstorming (20)                                 & problem understanding~(14)                    \\
                                                             & project requirement~(2)                         & to create storyboards~(1)                     \\
                                                             & diagram generation~(2)                           & visualizing user scenarios~(1)                \\
                                                             & architecture definition~(1)                     & user interview question generation (1) \\
                                                             & GUI mockups~(1) & design a framework template (1)               \\
                                                             & use case generation~(1)                                 &                        \\
{Software Development  \\\& Software Quality Assurance        } & information retrieval (55)                      & code generation (45)                          \\
                                                             & conceptual understanding (32)                   & code understanding (28)                       \\
                                                             & documentation generation (9)                    & to correcting syntax (8)                      \\
                                                             & unit test generation (6)                        & test case generation (5)                      \\
                                                             & to generate regular expression~(1)              & data test generation (2)                      \\
                                                             & game content generation (1)                     & image generation~(1)                          \\
                                                             & generating fake data for prototypes~(1)         & code comment generation~(1)                   \\
                                                             & regression testing generation~(1)               & validate JSON~(1)                             \\
                                                             & commit message generation (1)                   & code translation~(1)                          \\
Software Maintenance                                        & debugging (49)                                  & code refactoring (10)                         \\
                                                             & code review (7)                                 & code analysis (4)                             \\
                                                             & rubber duck debugging (2)                       & data analysis~(1)                             
\end{tblr}
\end{table}

\begin{figure}
    \centering
    \includegraphics[width=\linewidth]{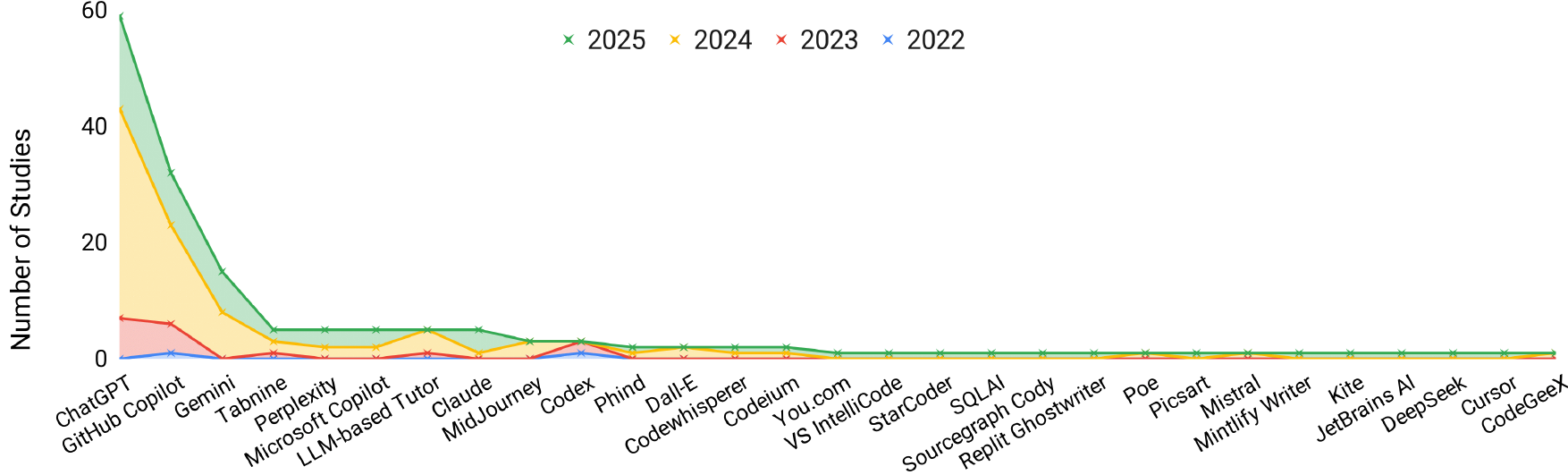}
    \caption{Distribution of LLM tools over years.}
    \label{fig:LLMToolsThroughoutYears}
\end{figure}

Figure \ref{fig:LLMToolsThroughoutYears} shows the distribution of 29 LLM tools in 73 studies. We found LLM tools designed to assist general activities (e.g., ChatGPT, Claude) and development (e.g., GitHub Copilot, Phind\footnote{\url{https://www.phind.com}}). We also found LLMs focused on image content generation (i.e., MidJourney and Dall-E). ChatGPT has been the most recurrent LLM tool in studies since 2023. We also identified an increase in the variety of LLM tools since 2024. Although all 73 studies mention proprietary LLMs, the open-source DeepSeek \cite{guo:2024, deng:2025} appeared for the first time in a publication in 2025. Whereas it is an arduous task to find alternative open-source LLMs that can challenge famous closed-source LLMs like ChatGPT and GitHub Copilot, Yang et al. \cite{zhou_yang:2024} explored and categorised the ecosystem of LLMs for coding tasks available in Hugging Face - the premier hub for transformer-based models. At the same time, we observe a gap in the literature regarding investigations comparing novice developers using open-source and proprietary LLMs. Ahmed et al. \cite{ahmed:2024} found that open-source LLMs can have different performance compared to closed-source LLMs depending on the programming language used. In this context, the study by Pereira et al. \cite{pereira:2024} appears as an initial comparison between open-source and proprietary in CS/SE Education, where the researchers developed a set of prompt examples to be used by CS students' prompts on ChatGPT and Mixtral\footnote{\url{https://mistral.ai/news/mixtral-of-experts}}, and LLaMA2\footnote{\url{https://www.llama.com/llama2}}, including many software development tasks. However, the authors argue about the need for further exploration with students to identify the benefits of open-source LLMs in SE education.  \label{rc:opensourceLLMsAndSEEducation}

\subsection{Requirement Engineering and Software Design.} \label{subsec:req}

This subsection focuses on how novice software developers utilise LLM tools for tasks associated with Requirements Engineering and Software Design, both of which require considerable creativity from developers \cite{jackson:2025, mohanani:2017}. It encompasses tasks such as brainstorming and problem understanding.

\subsubsection{Brainstorming} As an old practice with different variations (e.g., visual brainstorming, reverse brainstorming), brainstorming is widely employed in software engineering \cite{shih:2011, canedo:2022}. Brainstorming is one of the practices to achieve software innovations \cite{niva:2023}. Many studies mention novice developers using LLM tools to brainstorm solutions  \citeP{P14:boucher2024, P23:yilmaz2023, P25:zastudil2023, P33:hou2024, P34:haindl2024, P40:yabaku2024, P42:tan2024, P43:sergeyuk2025, P45:johansen2024, P48:adnin2025, P49:bikanga2024, P51:ramirez2025, P57:manley2024, P59:gorson2025, P61:alpizar2025, P62:choudhuri2025, P69:akhoroz2025, P71:lepp2025, P73:korpimies2024, P77:ouaazki2024}. For instance, Boucher et al. \citeP{P14:boucher2024} mention novice developers being encouraged to experiment with LLM tools during a summer internship program in Game Development. However, many discontinue using it for brainstorming after the trial. Given its inherent difficulty, it is not surprising that brainstorming is a recurring task in which novice developers use LLM tools. However, Jackson et al. \cite{jackson:2025} warn that overreliance on LLMs for creative activities might result in losing the \textit{creative intuition}. The findings of the experiment conducted by Kosmyna et al. \cite{kosmyna:2025} indicate that frequent AI tool users may experience skill atrophy in brainstorming. In education settings, educators should be aware of this collateral effect in the case that novice developers use LLM tools to brainstorm. In industry settings, developers could be affected by productivity expectations and oversight of these collateral effects, hindering their skills in the long term \cite{kam:2025}. It is the responsibility of software team leaders to see more than productivity metrics and ensure the evolution of their team.

\subsubsection{Problem Understanding} Accurate comprehension of the software requirement is a key cornerstone to achieving success in developing high-quality software solutions that satisfy stakeholders' needs \cite{werner:2020, bittner:2013}. Problem understanding consists of the first step during the problem-solving process, in which programmers interpret and clarify their understanding about the problem \cite{loksa:2016}. Many studies mention novice developers using LLMs to assist with problem understanding \citeP{P18:prather2024, P19:scholl2024, P28:margulieux2024, P38:waseem2024, P39:li2024, P44:shynkar2023, P48:adnin2025, P58:choi2024, P61:alpizar2025, P63:borghoff2025, P64:clarke2025, P67:alves2024, P69:akhoroz2025, P71:lepp2025}. We observe that all studies mentioning problem understanding involve CS/SE students as study participants. Those situations involve well-defined coding challenges, such as the "More Positive or Negative" coding challenge in the study by Prather et al. \citeP{P18:prather2024}, which is straightforward to pass all context to the LLM. In real-world scenarios, software developers need to build up the full picture by, for example, breaking the problem into smaller problems, conducting interviews, and asking for further clarification based on their understanding of the domain. With this in mind, LLMs present a potential for industry early-career software to quickly improve their domain expertise (e.g., Finance, Retail), and to develop efficient solutions. However, this potential remains unclear and requires verification. \label{rc:domainExpertise}

\subsubsection{Others.} The variety of possibilities in designing software architecture might make it a challenging task even for experienced developers \cite{capilla:2020}. Surprisingly, we only identified three studies reporting novice developers using LLMs for architecture definition \citeP{P38:waseem2024} and diagram generation \citeP{P29:xue2024, P63:borghoff2025} (UML, sequence diagram, flow chart). Even though a considerable percentage of the study participants is composed of CS/SE students working on small projects, we perceive the small number of papers as a research gap. \label{rc:architectureDefinition}

\subsection{Software Development and Software Quality Assurance.}

For tasks associated with Software Development and Software Quality Assurance, we found information retrieval, code generation, conceptual understanding, code understanding, and testing. We provide more details in the following paragraphs.

\subsubsection{Information Retrieval} Software developers frequently search the web for software resources, such as documentation and code examples, to support software development activities \cite{hora:2021, sim:2011}, complementing their lack of knowledge \cite{chen:2016}. They can spend up to 20\% of their time navigating between web pages \cite{hora:2021, niu:2017, brandt:2009}. 68.7\% of the primary studies mention novice developers using LLM tools for information retrieval related tasks (e.g., \citeP{P35:vasiliniuc2023, P17:prather2023, P18:prather2024, P19:scholl2024, P20:perry2023, P21:ross2023, P22:vaithilingam2022, P23:yilmaz2023}). For instance, Vasiliniuc et al. \citeP{P35:vasiliniuc2023} report novice developers using LLMs for understanding best practices, discovering new libraries, exploring trade-offs between different libraries, and finding links to relevant tutorials for in-depth learning. In this sense, LLMs appear to work by speeding up the information retrieval process, since developers would not need to browse many pages. However, it also disincentivises content creators from producing relevant content, as they may not be acknowledged or receive the monetising reward through ads \cite{fenwick:2024}. We recommend future studies to explore the perspective of tech content creators (e.g., Dev.to\footnote{\url{https://dev.to}} and Medium\footnote{\url{https://medium.com}}), focusing on achieving a solution that balances the interests of content creators and AI companies. \label{rc:balacingInterests}

\subsubsection{Code Generation} Coding is perceived by many developers as a challenging task \cite{becker:2023}. For this reason, code suggestion (e.g., code completion and code generation) is a relevant research topic in the software engineering community \cite{chen:2024}. AI-code suggestions can save developers' effort by providing personalised code snippets, in comparison with Stack Overflow. Surprisingly, only 56.2\% of the primary studies mention novice developers using LLM tools for code generation (e.g., \citeP{P7:prather2023, P8:russo2024, P9:barke2023, P11:rogers2024,P12:wang2024,P14:boucher2024,P15:weber2024,P16:ziegler2024,P17:prather2023,P18:prather2024,P19:scholl2024,P20:perry2023,P21:ross2023,P22:vaithilingam2022}). We believe the percentage of studies in educational settings influences this value, where CS/SE students would be restricted from using LLMs for code generation. At the same time, although using LLMs for code generation seems positive from the perspective of productivity, there is also the potential for impact on developers' code mental model \cite{liang:2025}. Developers' code mental model is developed and changed when developers work on the code base \cite{latoza:2006}. For early-career junior developers, they initially might struggle with a lack of knowledge involving the code base, but in the long term, that situation would change. We suggest a deep exploration involving the consequences of industrial junior developers using LLMs during companies' onboarding. \label{rc:companiesOnboard}

The study by Nguyen et al. \cite{nguyen:2022} demonstrated that the correctness of AI-based code varies according to the programming language. Educators should teach novice developers the flaws of LLM tools by also making them work with less prominent programming languages (e.g., Golang, Rust, and Kotlin). There is also the risk that LLMs influence the generation of monocultures \cite{jackson:2025, wu:2024, wenger:2025}. Novice software developers should be aware that there is no \textit{silver bullet} regarding a framework or programming language. Thus, there is a research gap regarding the extent to which LLMs influence the generation of monocultures in the population of novice developers, who may be more susceptible to overrelying on LLMs. \label{rc:monoculture}

\subsubsection{Conceptual Understanding} The foundation of programming extends from basic concepts (e.g., conditionals, looping) to advanced concepts (e.g., design patterns). Among the difficulties faced by CS students identified during their SLR, Qian et al. \cite{qian:2017} found difficulty in understanding object-oriented programming concepts. We identified 40\% of the primary studies mentioned novice developers using LLM tools for conceptual understanding. This highlights the educational potential held by LLMs \cite{wang:2024}, supporting novice developers in increasing their understanding of CS/SE concepts and practices. In an industrial context, the fast pace can make it unproductive to ask colleagues for help. However, in an educational context, mentorship interactions have a significant impact on novice developers. In summary, the impact of novice developers adopting LLM tools and their interaction with their mentors remains to be verified. \label{rc:mentorship}

\subsubsection{Code Understanding} Working with unfamiliar code is not an unusual scenario faced by developers \cite{taylor:2022, dagenais:2010}. However, when seeking information in the documentation, developers might find its content incomplete, outdated, or incorrect \cite{aghajani:2020}. This is why understanding code, or program comprehension, when the scale is the entire programming \cite{roehm:2012} - is essential. Qian et al. \cite{qian:2017} argue that understanding how code works may be difficult for novice developers. Not surprisingly, 35\% of the primary studies report novice developers using LLMs for code understanding. For instance, novice developers in the study by Tabarsi et al.\citeP{P53:tabarsi2025} rely on LLMs as a first option while trying to understand code, by copying and pasting the code snippet into ChatGPT. As a potential negative effect, the code-reading skill might not be developed. We suggest that researchers investigate the consequences of LLM adoption on code reading skills.  \label{rc:readingSkills}

\subsubsection{Testing} Software testing is a vital process to ensure quality and reliability of software systems \cite{wang:2024}, which are essential to the success of every product \cite{basili:2006}. Software testing consists of many tasks, such as test plan, test case preparation, and unit testing preparation \cite{whittaker:2002, wang:2024}. We found novice developers employing LLM tools for unit test generation \citeP{P2:qian2024, P8:russo2024, P21:ross2023, P31:sandhaus2024, P42:tan2024, P43:sergeyuk2025}, test case generation \citeP{P19:scholl2024, P43:sergeyuk2025, P59:gorson2025, P65:rasnayaka2024, P69:akhoroz2025}, data test generation \citeP{P17:prather2023, P43:sergeyuk2025}, and regression testing \citeP{P43:sergeyuk2025}. Wang et al. \cite{wang:2024} highlight the gap in understanding the capabilities of LLMs in solving software testing problems. At the same time, they found a successful example in the literature combining LLMs with traditional software techniques (e.g., mutation testing, differential testing). For this reason, we recommend that researchers investigate the effects of novice developers combining LLMs and traditional methods.

\subsubsection{Others} We also identified novice developers using LLMs for correcting programming language syntax \citeP{P19:scholl2024, P28:margulieux2024, P53:tabarsi2025, P64:clarke2025, P69:akhoroz2025, P73:korpimies2024, P75:ghimire2024, P78:lyu2025}. For novice developers who rely on LLMs for this assistance, they are losing the opportunity to familiarise themselves with the programming language and getting stuck in a state of unfamiliarity with syntax. In industry, there are many situations, like job interviews, where developers might not get external help while coding \cite{bell:2025, kaatz:2014}. We suggest that future research investigate the impact of the performance of LLM users in tech job interviews. There is also a gap in understanding how those LLM users would experience the transition to other programming languages. \label{rc:techJobInterview} \label{rc:learnOtherProgLang}

\subsection{Software Maintenance}

In this section, we discuss the tasks related to Software Maintenance, in which novice developers are using LLMs. It includes tasks such as debugging, code refactoring, and code review.

\subsubsection{Debugging} It includes detecting, locating, and correcting errors in a software \cite{layman:2013}. Traditionally, software developers seek support on online forums, such as Stack Overflow\footnote{\url{https://stackoverflow.com}}, as a common debugging approach \cite{chatterjee:2020, li:2013, mamykina:2011}. But, even with this support, the debugging process is still a challenging task, especially for novice developers \cite{li:2022, becker:2019}. Not surprisingly, we found 61.2\% of the studies reporting novice developers utilising LLM tools for debugging. Novice developers should be aware of limitations regarding LLMs, especially for debugging. From their experimental study, Majdoub et al. \cite{majdoub:2024} found DeepSeek scoring only around 65\%.  We were surprised to find two studies reporting novice developers using LLMs for \textit{rubber duck debugging}, which is an effective approach to identify the cause of a problem by verbalising how the code works \cite{whalley:2023, thomas:2019}. The literature lacks research on emerging and unique uses of LLMs such as this. \label{rc:uniqueUseCase}

\subsubsection{Code Refactoring} This process involves adjusting the software structure without changing its behaviour \cite{murphy:2011}. This practice helps contribute to enhancing code maintainability and reliability through, for example, removing code duplication and adoption of design patterns \cite{silvio:2016}. We only found 12.5\% of the primary studies reporting novice developers using LLMs for code refactoring \citeP{P8:russo2024, P18:prather2024, P43:sergeyuk2025, P54:zviel2024, P56:mailach2025, P64:clarke2025, P68:mendes2024, P69:akhoroz2025, P73:korpimies2024, P79:simaremare2024}. We believe there are two potential reasons: i) code refactoring is not widely employed by novice developers; ii) they might prefer to use traditional code refactoring tools (e.g., SonarQube\footnote{\url{http://www.sonarqube.org}}). Further investigation is required to provide an in-depth explanation.   \label{rc:codeReview}

\subsubsection{Code Review} This practice is well-known for its potential to improve the quality of software projects \cite{sadowski:2018, ackerman:1989, ackerman:1984}. Usually, code review is conducted by a developer who is not responsible for writing the code under review \cite{kononenko:2016}. We found a few studies mention novice developers utilising LLMs for code review \citeP{P34:haindl2024, P48:adnin2025, P58:choi2024, P63:borghoff2025, P65:rasnayaka2024, P67:alves2024, P80:alami2025}. Sadowski et al. \cite{sadowski:2018} argue that many companies adopted a lightweight code review process focusing on accelerating development. In that sense, LLMs can support novice developers by providing a preliminary code review.

\subsubsection{Others} We also found novice developers using LLMs for data analysis of user feedback or test results \citeP{P31:sandhaus2024}. They also use LLMs to perform code analysis \citeP{P5:styve2024, P19:scholl2024, P54:zviel2024, P65:rasnayaka2024}, focusing on performance improvements. While LLMs provide potential for code analysis, novice developers should also employ traditional static and runtime code analysis tools, such as FindBugs \footnote{\url{http://findbugs.sourceforge.net}} and ESLint\footnote{\url{https://eslint.org}}.

\begin{tcolorbox}[float=htpb!,colback=gray!5!white,colframe=gray!75!black,title=RQ2. What key software development tasks are novice developers using LLM-based tools for?]
We found evidence of novice developers using LLMs across all SE activities in the software development life cycle, except in Software Project Management. Most of the software development tasks are in Software Development and Software Quality Assurance (e.g., code generation, conceptual understanding). ChatGPT and GitHub Copilot are the most recurrent LLM tools in the studies. 
\end{tcolorbox}

\section{RQ3: What are the perceptions of novice software developers on using LLM-based tools?} \label{sec:rq3}

This section contains the findings regarding the third research question. We first present an overview of novice developers' perceptions in the primary studies. In the following subsections, we present the advantages, challenges, and recommendations reported by novice developers. 

\begin{figure}[ht]
    \centering
    \begin{subfigure}{0.45\textwidth}
        \centering
        \includegraphics[width=\textwidth]{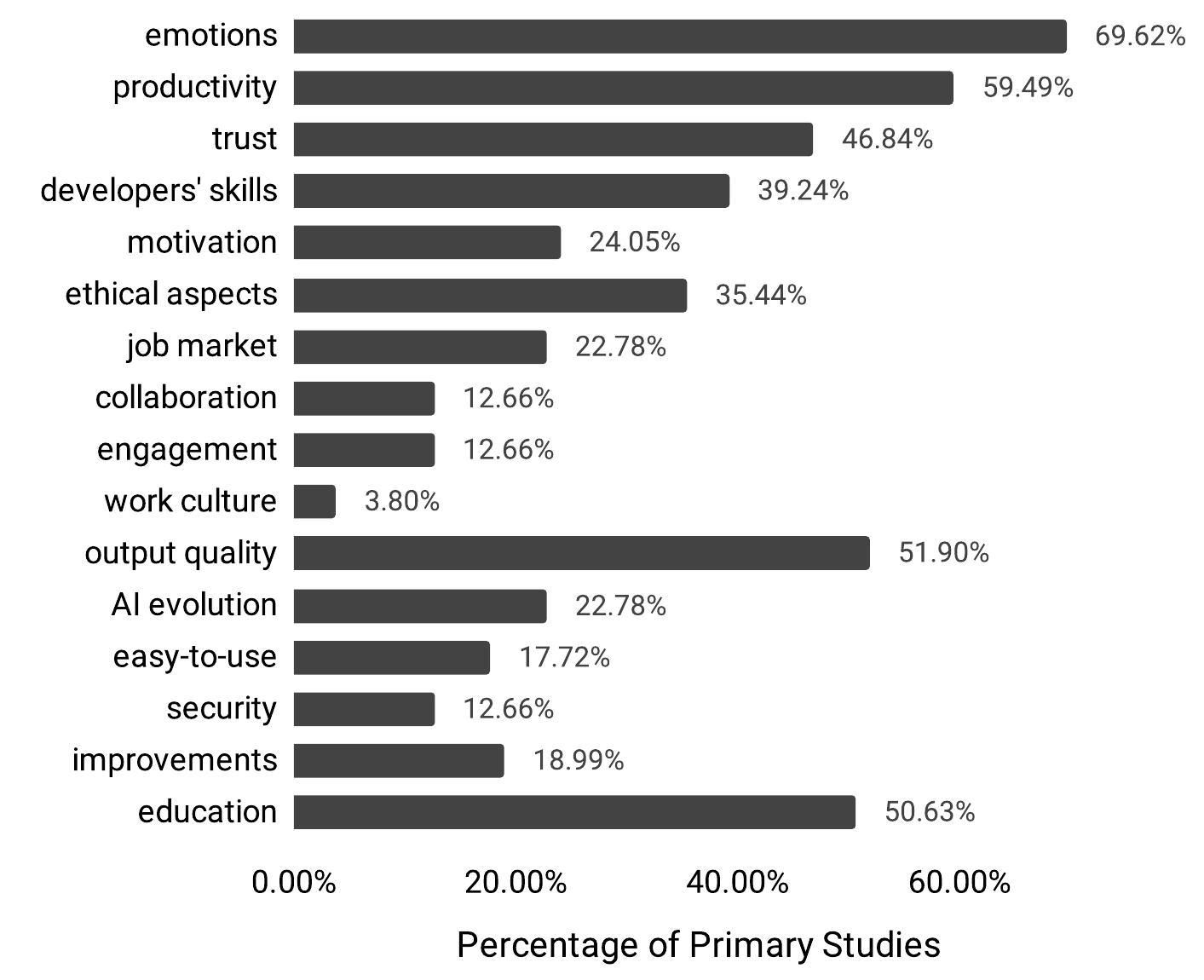}
        \caption{Distribution of the topics across primary studies.}
        \label{fig:perceptions}
    \end{subfigure}
    \hfill
    \begin{subfigure}{0.45\textwidth}
        \centering
        \includegraphics[width=\textwidth]{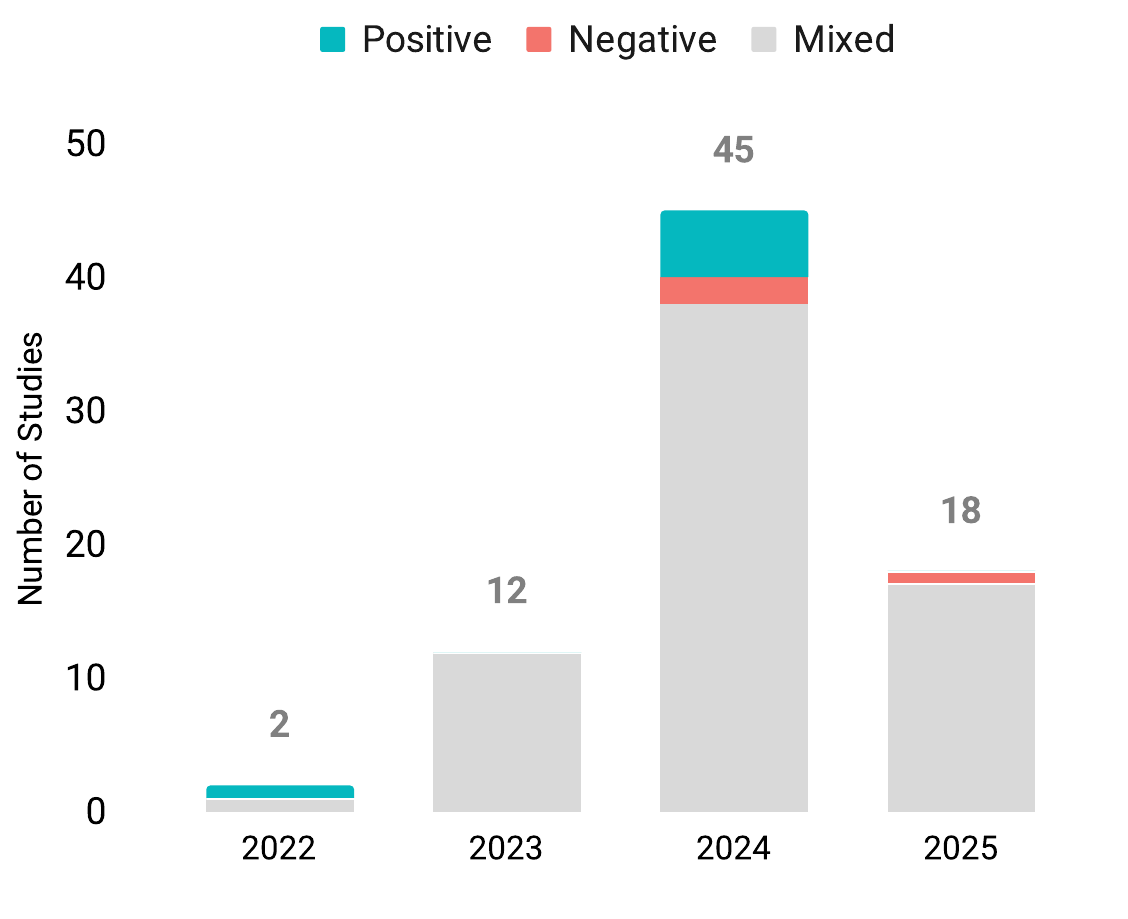} 
        \caption{Distribution of perceived impact of LLMs over the years.}
        \label{fig:impactPerceived}
    \end{subfigure}
    \caption{Overview of novice developers' perceptions of LLM adoption across 79 primary studies.}
\end{figure}

During our analysis, we classified the perceptions of novice developers into four categories: \textit{individual aspects}, \textit{industry collaborative aspects},  \textit{LLM-related aspects}, and educational aspects. Individual aspects consist of how LLM influences developers individually:  emotions, productivity, trust,  developers' skills, and motivation. Industry collaboration aspects refer to software developers working in teams, in the IT industry, using LLMs: ethical aspects such as privacy, copyright, and fairness/bias, job market, collaboration, engagement, and work culture. LLM aspects refer to non-functional and functional LLM features: output quality, AI evolution, user feedback regarding improvements, ease of use, and security. Education-related aspects refer to perceptions involving embracing LLMs in CS Education. Figure \ref{fig:perceptions} shows the distribution of the study participants' perceptions over 79 primary studies. Most studies report participants' emotions (e.g., fear, satisfaction, surprise, pessimism, frustration) towards LLMs and self-reported productivity. We believe that since LLM4SE emerged recently as a research topic, initial research has focused on understanding novice developers' general perception towards LLMs. Researchers could focus on investigating underexplored topics, such as the impact on work culture. \label{rc:underploredTopics}

\begin{figure}[ht]
    \centering
    \includegraphics[width=\linewidth]{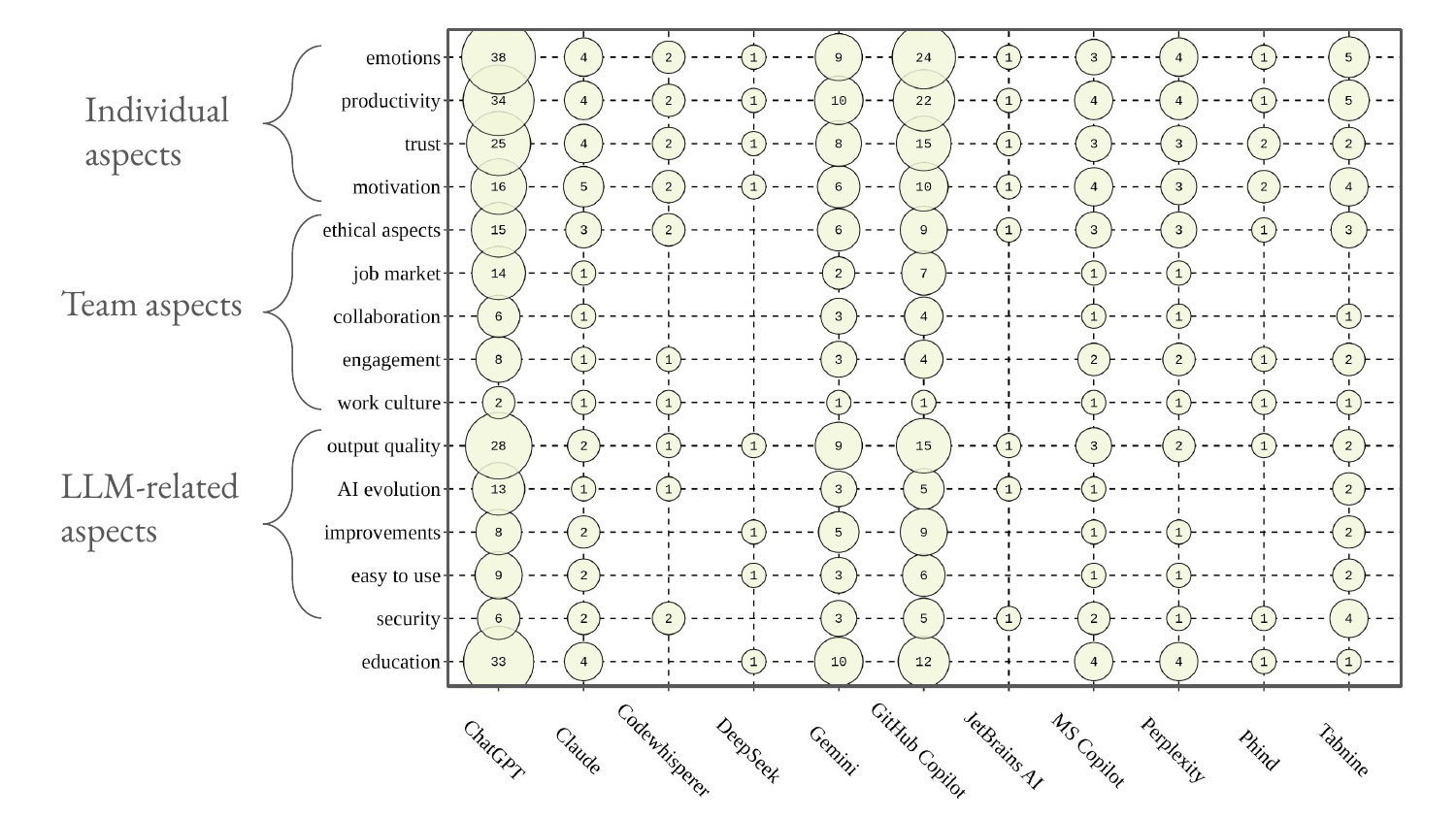}
    \caption{Distribution of topics discussed by novice developers across most recurrent LLM tools.}
    \label{fig:recurrentllmToolsPerceptions}
\end{figure}

Figure \ref{fig:impactPerceived} shows the distribution of the study participants' perceived impact of LLM adoption in terms of positive, negative, or mixed. Most studies describe a mix of feelings towards LLMs. However, we also observed a potential negative trend towards LLMs arising from 2024.  Figure \ref{fig:recurrentllmToolsPerceptions} shows the distribution of study participants' perceptions across the most recurrent LLM tools. Perceptions involving the impact of LLM adoption on self-report productivity, such as automating repetitive and tedious tasks \citeP{P3:amoozadeh2024}, and emotions (e.g., satisfaction and fear) appear commonly together. We observed a research gap in investigations of potential co-variables, such as work culture and satisfaction, influencing LLM adoption. \label{rc:coVariables}

% \textbf{Emotions.}    % % One of the participants in the study by Li et al. \citeP{P39:li2024} share their perception involving being open about LLM adoption: "The only risk you’re getting when you’re telling someone that you’re using AI tool(s), is that he's going to request you to show a demo because they also want to use it." % % However, Terragni et al. \cite{terragni:2024} argues that software developers are far away of being replaced by prompt engineers.

\subsection{RQ3a. What are the perceived and experienced advantages  of novice software developers on using LLM-based tools?} 

We found novice developers reporting the benefits of LLM adoption in 65 primary studies. We grouped them into four main categories: \textit{gains in productivity and efficiency}, \textit{learning opportunities}, \textit{additional assistance}, and \textit{improvement in code quality}. In most of the primary studies, study participants mention gains in productivity and efficiency. Despite this, LLMs show great potential for novice developers from the perspective of learning opportunities and additional assistance.

\textbf{Gains in Productivity \& Efficiency.} We found 45 studies in which study participants self-report the benefits related to improvements in productivity and efficiency. For example, LLMs can generate files in seconds \citeP{P1:mbizo2024, P5:styve2024, P38:waseem2024}, making novice developers complete tasks faster while spending less mental effort and time searching for information \citeP{P2:qian2024, P4:kazemitabaar2023, P11:rogers2024, P17:prather2023, P19:scholl2024, P21:ross2023, P23:yilmaz2023, P25:zastudil2023, P30:nam2024, P44:shynkar2023, P61:alpizar2025, P66:cipriano2024, P71:lepp2025}, typing \citeP{P14:boucher2024, P75:ghimire2024}, and troubleshooting \citeP{P31:sandhaus2024, P39:li2024, P44:shynkar2023}. Novice developers can stay focused and avoid distractions \citeP{P21:ross2023, P21:ross2023, P25:zastudil2023, P64:clarke2025, P72:hou2025}. LLMs were seen to automate many repetitive and tedious tasks  \citeP{P3:amoozadeh2024, P7:prather2023, P8:russo2024, P9:barke2023, P17:prather2023, P54:zviel2024, P63:borghoff2025, P65:rasnayaka2024}, speed up problem solving \citeP{P18:prather2024}, and reduce the effort to get started \citeP{P15:weber2024, P22:vaithilingam2022, P28:margulieux2024, P77:ouaazki2024}. For instance, novice developers can let Copilot generate boilerplate code composed of constructors and simple methods, saving their time \citeP{P5:styve2024, P17:prather2023, P24:jaworski2023, P55:kudriavtseva2025}. They also mentioned that they can generate code that may require minor changes \citeP{P30:nam2024, P32:haque2025}, with minor or no latency on feedback \citeP{P33:hou2024, P54:zviel2024, P69:akhoroz2025, P74:pang2024, P76:aruleba2023}.  Thus, novice developers can outsource tasks to LLMs \citeP{P7:prather2023}.  In this sense, we observe a research gap to extend what tasks novice developers would be comfortable outsourcing to LLMs. We suggest that researchers replicate the study by Masood et al. \cite{masood:2022} in the context of LLMs. \label{rc:outsourcing}

\textbf{Learning Opportunities.} We identified educational-related benefits in 25 studies. LLMs can assist novice developers to learn new coding approaches \citeP{P1:mbizo2024, P6:alkhayat2024, P18:prather2024, P19:scholl2024, P23:yilmaz2023, P25:zastudil2023, P64:clarke2025}, concepts \citeP{P3:amoozadeh2024, P29:xue2024, P47:aviv2024, P64:clarke2025}, and programming language syntax \citeP{P10:chen2024, P21:ross2023, P62:choudhuri2025, P69:akhoroz2025}. LLMs can also be used as a tutor \citeP{P19:scholl2024, P22:vaithilingam2022, P27:liu2024} and can speed up the learning process and reducing boredom \citeP{P41:boguslawski2024} via personalised learning \citeP{P51:ramirez2025, P62:choudhuri2025}. Novice developers can use LLMs to generate examples \citeP{P59:gorson2025, P61:alpizar2025}. They can also learn from the comments in AI-generated code \citeP{P76:aruleba2023}. LLMs were mentioned as having a latent potential for educational applications \citeP{P49:bikanga2024, P52:shao2025}, supporting self-learning in novice developers \citeP{P76:aruleba2023, P53:tabarsi2025}. However, the extension of the potential of instigating self-learning in early-career junior software developers, as well as supporting them to adjust to industry standards and practices, remains to be verified. \label{rc:selfLearning}

\textbf{Additional Assistance}. We identified 15 studies mentioning the potential of LLMs to provide assistance to novice developers. For instance, novice developers can ask \textit{dumb} questions without worrying about other's judgment \citeP{P19:scholl2024, P71:lepp2025} or bothering colleagues \citeP{P1:mbizo2024, P4:kazemitabaar2023, P10:chen2024, P11:rogers2024, P32:haque2025, P33:hou2024, P57:manley2024}. The constant availability of LLMs also stands out to novice developers \citeP{P11:rogers2024, P71:lepp2025}. LLMs also support novice developers to understand the reason behind errors \citeP{P5:styve2024} and during novel activities \citeP{P9:barke2023}, guiding them to a solution \citeP{P34:haindl2024}. LLMs can also support pair programming between novice developers \citeP{P53:tabarsi2025, P77:ouaazki2024}. While analysing the literature, we observed a research gap exploring how LLMs influence novice developers to overcome \textit{impostor phenomenon}. This occurs when individuals have an intense fear that others will perceive them as less capable than they actually are \cite{clance:1978}. Novice developers may suffer impostor phenomenon during transition from academia to industry, as highlighted by one of the participants in the study by Zhao et al. \cite{zhao:2024}: \textit{"I really want to be good enough for my current job...  I don't know if I can actually win or become an engineer after all... But now I have mainly high anxiety since I started [this] job... while it appears those who started with me are managing it well and getting stronger"}. Although the current literature has shown emerging studies focusing on software developers (e.g., \cite{guenes:2024, sun:2025}), Chen et al. \cite{chen:2024} point out the limitation in understanding it from a novice developer's perspective, especially in the context of LLMs. \label{rc:impostor}

\textbf{Improvement in Code Quality.} We found nine studies highlighting potential gains in code quality. LLMs can provide hints of potential improvements on code \citeP{P5:styve2024,P40:yabaku2024}, such as refactoring to become SOLID \citeP{P8:russo2024} and more organised \citeP{P29:xue2024}. They can also be used to identify minor errors in the code \citeP{P7:prather2023, P19:scholl2024, P77:ouaazki2024}. In summary, we found a small number of references to LLMs improving novice developers' code quality. We believe that this occurs because software developers usually focus on making the code functional, while putting code quality aside.  Techapalokul et al. \cite{techapalokul:2017} found that CS students maintain an attitude toward software quality even after having gained experience. They recommend that educators teach software quality concepts alongside the fundamentals of computing to convince students regarding the importance of software quality. We see the potential of educators using LLMs to generate sophisticated examples involving \textit{good code} and \textit{bad code} when teaching programming. Future research could explore how the adoption of LLM in education can improve CS/SE students' code quality. \label{rc:codeQuality}

\subsection{RQ3b. What are the perceived and experienced  challenges \& limitations faced by novice software developers while using LLM-based tools?} 
    
We identified novice developers reporting challenges and limitations associated with LLM adoption in 68 primary studies. We grouped them into three main categories: \textit{novice not ready for LLMs}, \textit{losing learning opportunities}, and (LLMs are) \textit{not a good fit for novice developers}. Most studies mention aspects that highlight how novice developers are not adequately prepared to handle the risks of using LLM tools.

\textbf{Novices are Not Ready for LLMs.} We found 35 primary studies reporting aspects showing that novice developers are not mature enough to use LLMs. AI suggestions are based on the trained dataset that might not reflect organisation realities or other niche topics, generating sub-optimal solutions \citeP{P1:mbizo2024, P2:qian2024, P3:amoozadeh2024, P8:russo2024, P46:choudhuri2024, P69:akhoroz2025}. Thus, those tools might generate low quality or incorrect solutions not following, for example, software engineering best practices \citeP{P6:alkhayat2024, P7:prather2023, P8:russo2024, P10:chen2024, P15:weber2024, P18:prather2024, P19:scholl2024, P21:ross2023, P23:yilmaz2023, P25:zastudil2023, P27:liu2024, P29:xue2024, P30:nam2024, P31:sandhaus2024, P32:haque2025, P33:hou2024, P39:li2024, P40:yabaku2024, P41:boguslawski2024, P43:sergeyuk2025, P45:johansen2024, P46:choudhuri2024, P47:aviv2024, P77:ouaazki2024}. Novice developers might find it difficult to evaluate the LLM suggestions, which apparently look right \citeP{P62:choudhuri2025}, due to a lack of background knowledge \citeP{P10:chen2024, P25:zastudil2023, P26:shoufan2023}. They might not know when to use LLMs \citeP{P60:rahe2025}, knowing that LLMs struggle with complex context \citeP{P47:aviv2024, P77:ouaazki2024} and might misunderstand users' prompts \citeP{P71:lepp2025}. To avoid this, novice developers would need to provide very specific and clear prompts to generate a desirable solution, but it is challenging to frame that specific prompt \citeP{P8:russo2024, P10:chen2024, P62:choudhuri2025, P63:borghoff2025}. And when faced with the scenario where the prompt lacks sufficient context \citeP{P19:scholl2024}, a novice might increase context \citeP{P78:lyu2025}. However, while increasing the context, the response time also increases \citeP{P77:ouaazki2024}. They would need to restart the conversation because the model loses context \citeP{P53:tabarsi2025} or get stuck in the same incorrect response \citeP{P19:scholl2024, P78:lyu2025}. They also might be arm-wrestling with LLMs \citeP{P19:scholl2024, P67:alves2024}, struggling because they do not have enough reasons to refute AI bad code suggestions.

\textbf{Losing Learning Opportunities.} We found educational-related challenges in 31 primary studies. By outsourcing tasks to LLMs, novice developers would miss essential learning experiences  \citeP{P1:mbizo2024, P9:barke2023, P10:chen2024, P11:rogers2024, P19:scholl2024, P25:zastudil2023, P26:shoufan2023, P29:xue2024, P32:haque2025, P35:vasiliniuc2023,  P36:yang2024, P37:keuning2024, P44:shynkar2023, P49:bikanga2024, P54:zviel2024, P55:kudriavtseva2025, P57:manley2024, P61:alpizar2025, P65:rasnayaka2024, P71:lepp2025, P79:simaremare2024}. For instance, ChatGPT might provide a full solution rather than a partial solution during learning \citeP{P57:manley2024}. Overreliance on LLMs might make a novice developer dependent \citeP{P2:qian2024, P23:yilmaz2023} and lazy \citeP{P23:yilmaz2023, P72:hou2025, P79:simaremare2024}, rather than eager to learn. This attitude could hinder their potential of mastering programming languages and concepts \citeP{P70:padiyath2024}. By just copying and pasting AI suggestions, they also lose the opportunity to understand the codebase \citeP{P7:prather2023}, which is a foundation to posterior debugging processes \citeP{P22:vaithilingam2022}. Novice developers might spend a lot of time creating very descriptive prompts but not understand the reason behind the AI response, similar to a \textit{blackbox} \citeP{P4:kazemitabaar2023, P25:zastudil2023}. There are also the potential negative implications on developers' confidence in their intuition \citeP{P2:qian2024}, where the developer would trust LLM suggestions over their own, disturbing their thinking process \citeP{P7:prather2023, P9:barke2023, P15:weber2024, P62:choudhuri2025}. We suggest future research in-depth explorations on the impact on developers' confidence, which might influence developers' productivity. The inconsistency of LLM suggestions \citeP{P19:scholl2024, P54:zviel2024}, by providing different responses to the same prompt, also provides a challenge for learning.

\textbf{Not a Good Fit for Novice Developers.} We identified 12 primary studies reporting aspects showing that LLMs are not recommended for novice developers. Developers would need to customise LLMs to avoid, for example, ChatGPT generating code with unnecessary comments \citeP{P76:aruleba2023}, or force it to match codebase style \citeP{P22:vaithilingam2022, P62:choudhuri2025}. However, novice developers might not know how to adjust complex parameters (e.g., temperature) \citeP{P49:bikanga2024}. There is a concern that novice developers would upload proprietary code into LLMs \citeP{P1:mbizo2024, P24:jaworski2023, P42:tan2024}. AI can be misleading, for example, through AI sycophancy, which is another trap novice developers could fall into \citeP{P6:alkhayat2024}. More than demonstrating friendliness, \textit{AI sycophancy} \cite{sun:2025, sharma:2023} makes LLMs align their responses with users' perspectives even though users provide incorrect views. In this sense, Bo et al. \cite{bo:2025} propose to adjust LLMs to act more actively to correct users' faulty assumptions. The practical applicability of such approaches remains to be verified. \label{rc:adjustingLLMsToActActively}

\textbf{Others.} Novice developers also mentioned that code generation takes away the best part of software development: coding \citeP{P65:rasnayaka2024}. LLM adoption may also lead to a loss of sense of ownership \citeP{P76:aruleba2023}. Bird et al. \cite{bird:2011} found that code ownership has a strong relationship with software quality. Thus, while adopting LLMs, there is a potential to increase the number of system failures. There is also a loss of mentorship, where incoming novice developers might not seek support from their seniors \citeP{P72:hou2025}, as well as the increasing expectations of employers \citeP{P17:prather2023}. We suggest future research survey software team leaders and IT managers to get their perceptions. In addition, when seeking to use high-performance models, they may require payment \citeP{P71:lepp2025}. \label{rc:expectations}

\subsection{RQ3c. What are the  recommendations \& best practices  suggested by novice software developers while using LLM-based tools?}

We found novice developers reporting the best practices regarding LLM adoption in 34 primary studies. We grouped them into three main categories:\textit{ prompt engineering}, \textit{cautious towards LLMs}, and \textit{when to use LLMs}. Most of the selected studies include recommendations to developers to be analytical towards AI suggestions. This shows that novice developers have a prudent attitude regarding the LLM adoption. Most of the recommendations compiled below are perhaps applicable to software developers at all levels of expertise, except those regarding when to use LLMs, which are more specific to novice developers.

\textbf{Prompt Engineering.} We identified study participants using a prompting engineering approach in ten studies. For instance, break prompts into small tasks \citeP{P9:barke2023, P10:chen2024, P53:tabarsi2025, P69:akhoroz2025}, provide context in prompt \citeP{P68:mendes2024}, asking for specific code segments \citeP{P14:boucher2024}. Study participants also suggest asking follow-up questions to guide LLM towards the solution \citeP{P26:shoufan2023, P32:haque2025, P36:yang2024}. Study participants also highlight the importance of learning prompting engineering practices \citeP{P54:zviel2024,P65:rasnayaka2024, P69:akhoroz2025}, and prompting in English as a strategy to improve the accuracy of LLMs \citeP{P54:zviel2024}. However, novice developers with English as a second language might face a language barrier \citeP{P79:simaremare2024}. These results indicate that there is an emerging understanding by novice developers about the importance of prompting engineering when using LLMs.

\textbf{Cautious towards LLMs.} We found that nineteen primary studies mentioned a cautious attitude towards LLMs. Potential negative consequences for misuse of LLMs include, for example, disruption of the developer's mental coding flow \citeP{P21:ross2023}. In this sense, study participants suggest that novice developers should adopt an analytical attitude towards LLMs, evaluating each AI suggestion \citeP{P2:qian2024, P18:prather2024, P33:hou2024, P52:shao2025, P64:clarke2025} and modifying it to their context. They recommend the adoption of tools from reliable AI vendors with a focus on security, and the use of external platforms (e.g., documentation, Stack Overflow, code scanning) to double-check AI suggestions before accepting them \citeP{P8:russo2024, P9:barke2023, P32:haque2025, P53:tabarsi2025, P55:kudriavtseva2025, P61:alpizar2025, P62:choudhuri2025, P68:mendes2024, P69:akhoroz2025}. They also recommend that novice developers treat AI-generated code as a baseline \citeP{P31:sandhaus2024, P64:clarke2025}, not requesting to generate the entire solution, because it may generate wrong code \citeP{P3:amoozadeh2024, P18:prather2024, P29:xue2024}. Study participants suggest novice developers collaborate with other developers to improve their understanding of those tools \citeP{P6:alkhayat2024}. There is also a recommendation to not use ChatGPT for system-wide testing due to security concerns \citeP{P53:tabarsi2025}, since LLMs struggle with complex tests.

\textbf{When to use LLMs.} We found nine studies mentioning recommendations to help novice developers decide when to use LLM tools. Study participants recommend novice developers only go after LLM tools when they fail to reach a solution by themselves \citeP{P17:prather2023, P36:yang2024, P60:rahe2025, P61:alpizar2025, P64:clarke2025, P74:pang2024}. LLMs can also be used to improve novice developers' code quality \citeP{P53:tabarsi2025}. Focus on improvements, participants of the study by Alkhayat et al. \citeP{P6:alkhayat2024} suggest that Virtual Reality developers familiarise themselves first with the Unity interface and the C\# programming language to be able to comprehend ChatGPT suggestions. In that sense, study participants also recommend novice developers use LLMs as a learning tool to support them in solidifying the basis \citeP{P69:akhoroz2025, P70:padiyath2024}.

\begin{tcolorbox}[float=htpb!,colback=gray!5!white,colframe=gray!75!black,title=RQ3. What are the perceptions of novice software developers on using LLM-based tools?]
A majority of the studies present a mix of positive and negative aspects related to novice developers adopting LLM tools. Gains in productivity appear as the most commonly perceived advantage, while many study participants in the primary studies recognise aspects where LLMs and novice developers are not a good fit. Regarding recommendations, most of the studies suggest a cautious approach towards LLMs. 
\end{tcolorbox}

\section{RQ4. What are the limitations and recommendations for future research that we can distil based on the primary studies?} \label{sec:rq4}

This section contains the findings regarding the fourth research question. We first present an overview of study limitations and future research needs suggested in the primary studies.

\subsection{Primary Study Limitations}

We categorised the primary study limitations into three main categories: \textit{limitations in approach}, \textit{limitations in data collection and analysis}, and \textit{limitations in findings}. Most studies report limitations in data collection and analysis that are not directly related to LLMs.

\textbf{Limitations in Approach.} We identified limitations regarding the study approach in thirteen primary studies. This includes limitations in the design of the task \citeP{P2:qian2024, P4:kazemitabaar2023, P12:wang2024, P13:gardella2024, P30:nam2024}, where confounding variables \citeP{P34:haindl2024, P36:yang2024, P37:keuning2024, P42:tan2024, P46:choudhuri2024}, such as the level of familiarity with AI tools, could impact the results. A few studies recognise the probabilistic nature of LLMs during experiments, difficulty in study replication \citeP{P15:weber2024, P66:cipriano2024}. Other studies acknowledge that the non-realistic experimental environment could impact the findings  \citeP{P7:prather2023, P15:weber2024, P20:perry2023, P30:nam2024, P36:yang2024, P76:aruleba2023, P78:lyu2025, P80:alami2025}. Another study acknowledged a potential deterioration in the performance of used LLMs because the study participants prompted the LLMs in German, their native language, instead of English \citeP{P60:rahe2025}. This understanding is consistent with the literature that demonstrates that ChatGPT shows poor performance for prompts in languages other than English \cite{zhang:2023,lai:2023}.

\textbf{Data Collection \& Data Analysis.} We identified limitations in data collection and analysis in 46 primary studies. For example, limitations in sampling (e.g., participants' geographic location, number of participants), which could not reflect the opinion of the entire SE population, are a common limitation to most empirical SE studies \citeP{P1:mbizo2024, P2:qian2024, P3:amoozadeh2024, P6:alkhayat2024, P8:russo2024, P9:barke2023, P10:chen2024, P13:gardella2024, P14:boucher2024, P16:ziegler2024, P18:prather2024, P20:perry2023, P25:zastudil2023, P29:xue2024, P30:nam2024, P32:haque2025, P33:hou2024, P34:haindl2024, P37:keuning2024, P38:waseem2024, P39:li2024, P41:boguslawski2024, P42:tan2024, P43:sergeyuk2025, P44:shynkar2023, P45:johansen2024, P46:choudhuri2024, P52:shao2025, P53:tabarsi2025, P55:kudriavtseva2025, P56:mailach2025, P58:choi2024, P62:choudhuri2025, P66:cipriano2024, P68:mendes2024, P71:lepp2025, P72:hou2025, P73:korpimies2024, P74:pang2024, P75:ghimire2024, P76:aruleba2023, P77:ouaazki2024, P78:lyu2025, P79:simaremare2024}. They also acknowledge potential self-selection bias where, for example, participants interested in the study topic would be more interested in joining their studies  \citeP{P10:chen2024, P37:keuning2024, P46:choudhuri2024, P55:kudriavtseva2025, P56:mailach2025, P60:rahe2025, P62:choudhuri2025}. Participants could also misinterpret researchers' questions \citeP{P32:haque2025, P34:haindl2024, P42:tan2024}. Regarding data analysis, researchers acknowledge potential research bias during the qualitative data analysis process \citeP{P59:gorson2025, P60:rahe2025, P73:korpimies2024}. All these reported study limitations are common in empirical studies.

\textbf{Limitations in Findings.} We identified 25 primary studies reporting limitations regarding findings. This includes self-report data \citeP{P1:mbizo2024, P3:amoozadeh2024, P19:scholl2024, P32:haque2025, P36:yang2024, P37:keuning2024, P49:bikanga2024, P52:shao2025, P54:zviel2024, P55:kudriavtseva2025, P58:choi2024, P61:alpizar2025, P69:akhoroz2025, P71:lepp2025, P72:hou2025, P73:korpimies2024} and memory bias \citeP{P3:amoozadeh2024, P32:haque2025, P42:tan2024, P62:choudhuri2025}. A few studies recognise that findings related to ChatGPT and Copilot could not be generalised to all LLM tools \citeP{P2:qian2024, P18:prather2024, P22:vaithilingam2022, P66:cipriano2024}. A few studies recognise that their findings do not converge over a long period of observation and experiment \citeP{P2:qian2024, P9:barke2023, P29:xue2024, P37:keuning2024, P77:ouaazki2024}. Only a few studies recognise that the fast pace of evolution of AI tools is making study findings outdate more quickly \citeP{P9:barke2023}. In that sense, we suggest that future research indicate LLM versions or features available when the study was conducted to enable readers to understand LLM capabilities during that period. 

\subsection{Future Research Needs}

We found 76 primary studies proposing future investigations. We grouped them into five main categories: \textit{exploratory studies}, \textit{development and improvement of guidelines and tools}, \textit{replication studies}, \textit{extension studies}, and \textit{longitudinal studies}. Most of the primary studies suggest topics for future exploratory studies.

\textbf{Exploratory studies.} We identified key recommendations for research explorations in 30 primary studies related to novice developers. For instance, Boucher et al. suggest that researchers explore to what extent LLM tools can support game development \citeP{P14:boucher2024}. They also recommend more studies exploring attitudes (e.g., resistance) in early career software professionals  \citeP{P14:boucher2024, P29:xue2024, P73:korpimies2024}, taking in consideration gender minority \citeP{P17:prather2023}, and communities and organisational settings  \citeP{P17:prather2023}. Investigate better approaches for developers to understand long blocks of generated code \citeP{P21:ross2023, P9:barke2023}, as well as ways to better control LLM tools \citeP{P23:yilmaz2023, P9:barke2023}. Future studies could explore how LLM tools influence pair programming \citeP{P79:simaremare2024} and novice developers' code reuse \citeP{P60:rahe2025}. There are recommendations for research focusing on organisational aspects (e.g., culture, size) and how they affect the adoption by software practitioners. \citeP{P39:li2024, P70:padiyath2024}. Studies also suggest research on how group dynamics in software development teams with developers with multiple levels of expertise are affected by LLM adoption \citeP{P15:weber2024, P39:li2024, P6:alkhayat2024, P33:hou2024}. In this context, one of the participants in the study by Kemel et al. \cite{kemell:2025}, which conducted an exploratory case study of 7 European companies, mentioned that: \textit{"For me, when I started, I had some difficulty in making some code contributions because the code base was so huge. And it took... a lot, a lot of time to get used to it. So if I, if I had [GitHub] Copilot back then, I think I would have made contributions much earlier"}. Based on this, we suggest future studies seek to understand how junior developers can leverage LLM tools to improve their understanding of large code bases.

In education, many studies suggest more investigations on how LLM tools can be incorporated into computer education, aligning with educational objectives and learning outcomes \citeP{P7:prather2023, P12:wang2024, P11:rogers2024, P23:yilmaz2023, P29:xue2024, P39:li2024, P44:shynkar2023, P46:choudhuri2024, P47:aviv2024, P64:clarke2025, P67:alves2024, P5:styve2024, P36:yang2024}, understanding which core skills expected of graduates \citeP{P13:gardella2024}. But also, focused on adjusting assessment to properly evaluate students in an LLM era, mitigating the risk of them using LLMs to cheat \citeP{P4:kazemitabaar2023, P6:alkhayat2024, P11:rogers2024}. They also suggest exploration the impact on novice developers' learning and motivation while being exposed to numerous LLM suggestions \citeP{P18:prather2024, P4:kazemitabaar2023, P3:amoozadeh2024, P26:shoufan2023}, how novice developers break down tasks and write prompts for LLMs \citeP{P4:kazemitabaar2023}, and comparison between traditional learning and AI-enhanced learning with student traditional learning, leveraged with personalised student assistance \citeP{P54:zviel2024, P46:choudhuri2024, P60:rahe2025}. In this context, there is also a suggestion for exploring the impact of LLM tools on individual versus team-based learning \citeP{P54:zviel2024}.

\textbf{Development and Improvement of Guidelines and Tools.} We identified ten primary studies discussing ideas for LLM tools and guidelines. For instance, studies suggest designing new metrics to improve the alignment of LLM tools with individual preferences \citeP{P20:perry2023, P23:yilmaz2023, P80:alami2025}. Regarding Computer Education, they suggest the development of specialised GPT models focused on educational aspects that could be incorporated into well-known educational environments \citeP{P19:scholl2024, P38:waseem2024, P56:mailach2025}, supporting feedback across submissions \citeP{P27:liu2024}. In this sense, ChatGPT provides a collection of customised GPT models\footnote{\url{https://chatgpt.com/gpts}} that CS/SE educators can use. Educators can also create their own customised GPT model by following the steps in the study by \cite{kabir:2025}, which created a neurosurgical research paper writer and medi research assistant. To support effective adoption of LLM in Computer Education, it is also necessary to develop an easy way to cite AI support in students' code \citeP{P57:manley2024}. Given the necessity to understand how LLM tools work, Barke et al. \citeP{P9:barke2023} suggest that the developer community work collaboratively to create a "community guide", including best prompts and comments used to achieve the best results. We found the subreddit \texttt{r/ChatGPTPromptGenius}\footnote{\url{https://www.reddit.com/r/ChatGPTPromptGenius}} that includes many relevant discussions that help to understand the behaviour of ChatGPT. Future studies could explore submissions on this subreddit, similar to when Kuutila et al. \cite{kuutila:2024} explored the subreddit \texttt{r/ProgrammerHumor}. To improve understanding of debugging, Akhoroz et al. \citeP{P69:akhoroz2025} suggest LLM tools to include visual explanation. One of the studies recommends the development of guidelines for the responsible use of LLMs in the industry \citeP{P45:johansen2024}. Berengueres et al. \cite{berengueres:2024} explored the ethical aspects related to how LLM service providers could regulate their LLM services, but a research gap remains regarding practical guidelines.

\textbf{Replication studies.} We found nine primary studies suggesting replication of their empirical studies. Replication of studies is a valuable strategy adopted by the research community for over 30 years, aiming to verify and compare the findings of previous studies \cite{da:2014, bezerra:2015}. Studies recommend validating findings in other educational settings \citeP{P26:shoufan2023, P51:ramirez2025, P59:gorson2025, P70:padiyath2024} and other areas \citeP{P19:scholl2024}. They also recommend that future studies update their work with recent LLM versions \citeP{P13:gardella2024, P33:hou2024}. For instance, the study by Gardella et al. \citeP{P13:gardella2024} was conducted using the GitHub Copilot version that did not include a chat mode for code generation, but only a comment-based code generation. 

\textbf{Extension studies.} We found twenty-six primary studies recommending future research to extend their studies. For instance, they suggest expanding the studies by incorporating other LLM tools \citeP{P2:qian2024, P18:prather2024, P49:bikanga2024, P54:zviel2024, P60:rahe2025}, different tasks \citeP{P13:gardella2024}, and different projects \citeP{P63:borghoff2025}, as well as more diversity and larger population of participants \citeP{P1:mbizo2024, P4:kazemitabaar2023, P10:chen2024, P13:gardella2024, P18:prather2024, P19:scholl2024, P24:jaworski2023, P30:nam2024, P33:hou2024, P34:haindl2024, P37:keuning2024, P41:boguslawski2024, P42:tan2024, P45:johansen2024, P53:tabarsi2025, P64:clarke2025, P68:mendes2024, P73:korpimies2024, P76:aruleba2023, P77:ouaazki2024}, and different mixed-method approaches \citeP{P2:qian2024}. This is consistent with the current landscape, where there is a greater diversity of LLM tools nowadays, such as Grok\footnote{\url{https://grok.com}} and Qwen\footnote{\url{https://qwen.ai}}, compared to 2022 when ChatGPT became recognised worldwide. They also suggest conducting studies with long-term LLM users \citeP{P15:weber2024, P24:jaworski2023} and real-world settings \citeP{P30:nam2024}.

\textbf{Longitudinal studies.} We identified fourteen primary studies recommending longitudinal research on the impact of LLM adoption in general, as well as on novice developers' learning \citeP{P13:gardella2024, P64:clarke2025, P75:ghimire2024}. This challenging approach, which involves data collection related to a long time frame, has been utilised by SE researchers for a long time (e.g., \cite{fitzgerald:1999, kemerer:2002}). While facing a massive amount of potential data, it is essential to efficiently select the most appropriate metrics to be collected. In our primary studies, authors suggest the following metrics: perceptions and challenges over time  \citeP{P1:mbizo2024, P7:prather2023, P23:yilmaz2023}, and adoption patterns and trends \citeP{P8:russo2024, P52:shao2025, P69:akhoroz2025}. It is also necessary to clearly delimit the scope, for example, by investigating the effects of an evidence-based software engineering training \cite{pizard:2022}. Studies suggest that future long-term investigations are required on the implications of adopting LLM tools in pair programming \citeP{P79:simaremare2024} and career readiness \citeP{P65:rasnayaka2024}. They also recommend long-term studies involving multiple organisations \citeP{P37:keuning2024} and real-world large software projects \citeP{P30:nam2024}, similar to Cedrim et al. \cite{cedrim:2017}, which conducted a longitudinal study on the impact of refactoring on code smells using 23 software projects. We observe that future research could contribute by building a large dataset that contains the interactions of novice developers with LLMs for exploration by the research community.

\begin{tcolorbox}[float=htpb!,colback=gray!5!white,colframe=gray!75!black,title=RQ4. What are the limitations and recommendations for future research that we can distil based on the primary studies?]
We identified limitations across three main categories: limitations in data collection and analysis (57.5\%), limitations in findings (31.2\%), and limitations in approach (16.2\%). We also identified the reported future research needs, classifying them into five main categories: exploratory study (37.5\%), extension study (32.5\%), longitudinal studies (17.5\%), development and improvement of guidelines and LLM tools (12.8\%), and replication study (11.25\%).
\end{tcolorbox}

\section{Discussion and Research Roadmap} \label{sec:discussion}

In this section, we discuss the SLR findings in terms of implications for software developers and educators. Then, we synthesise the research directions for future research.

\textbf{Implications for Practice.} We discussed in section \ref{sec:rq2} the reported SE tasks that novice developers are using LLMs. Most of the SE tasks go under software development and software quality assurance activities. In the context of novice developers, there is a greater potential for using LLMs for concept understanding, which would improve their skills in the long term. At the same time, there are many recommendations for novice developers to act with caution, especially when they are consolidating the fundamental CS concepts. In this sense, the key strategy is to understand when to use LLMs (e.g., after having gained familiarity with the programming language syntax) and how to use LLMs (i.e., prompt engineering). Software team leaders have the responsibility of looking after their team members, alerting them about the potential LLM pitfalls (e.g., introducing bugs or non-functional code). They are also responsible for creating an environment that embraces team collaboration, vital to the success of software development \cite{strode:2022}. According to Strode et al. \cite{strode:2022}, \textit{Working in personal caves} hinders team effectiveness by potentially affecting the common understanding of the process. By overrelying on LLMs, novice developers could underappreciate team collaboration. Software team leaders should monitor those developers and promote team spirit.

\textbf{Implications for Education.} Given the potential consequences (e.g., disruption of the system in the production environment due to AI-generated code introducing errors), novice developers should be clearly instructed about the limitations of using LLMs. We suggest that educators present the limitations of LLMs, for example, how their performance depends on the context (e.g., programming language, task). For this reason, educators could incorporate realistic scenarios and projects that students would face in industry using role-play and simulations. For instance, a CS/SE student would act as employees from a company in a domain very concerned about privacy and data protection, such as finance and government. In this context, Zheng et al. \cite{zheng:2024} conducted an evaluation of different LLM tools, using a dataset that includes 12 application domains. They reported that famous LLMs often perform poorly in domain-specific code generation tasks. Another example, the students could work on legacy projects in less prominent programming languages (e.g., Ruby, PHP) or programming languages in which popular LLMs perform poorly.

\subsection{Directions for Future Work}

In previous sections, we have indicated several research gaps while presenting the findings. We synthesise all these research opportunities in Table \ref{tab:recommendations}. In the rest of this section, we present an additional set of recommendations to the SE research community for future exploration.

\textbf{Novices \& Vibe Coding.} We are surprised that vibe coding did not appear in the publications in 2025,  even though there is a large repercussion in the software industry \cite{harkar:vibecoding}. The term \textit{vibe coding} refers to an emerging new programming style where developers completely rely on LLM tools to generate good quality code using natural language, while developers disconnect from activities directly related to code (e.g., code writing, code reading) \cite{sarkar:2025}. Based on the potential of vibe coding, there are discussions on Reddit (e.g., \cite{reddit:2025, reddit2:2025}) exploring the implications of vibe coding to software developers. From a novice developer's perspective, vibe coding reinforces the idea of AI replacing software developers. However, experienced developers understand that code development is only one of the phases of the software development life cycle. There is also the hidden cost (e.g., power and water consumption) of using LLM tools \cite{shi:2025}. In this sense, we recommend investigations regarding the impact of vibe coding on novice developers.

\textbf{LLMs for Software Project Management.} As mentioned in Section \ref{sec:rq2}, there is a research gap in software management. Software management encompasses practices to supervise activities across the software development life cycle \cite{mills:1980}. Zhang et al. \cite{zhang:2023} identified during their systematic survey the potential of project managers to employ LLMs to analyse developers'  emotions \cite{imran:2024}, such as \textit{frustration} due to postponing merging pull requests. Previous studies explored the connection between developers' emotions and their productivity \cite{murgia:2014, crawford:2014, wrobel:2013}. With this in mind, we recommend that researchers explore the extent to which LLMs can improve project managers' support for early-career developers.   In this context, we also observe a research gap in studies focused on the impact of LLM tools on the mental health of novice developers. Our suggestion aligns with a suggestion in the position study by Meem \cite{meem:2024} that proposes that researchers investigate what features of LLM tools can be helpful or harmful for the mental health of software practitioners. Another motivation for this research includes the findings from the study by Graziotin et al. \cite{graziotin:2017, graziotin:2018}, which identified from their survey with 181 developers that mental disorders (e.g., anxiety, depression) cause delays in the software development process.

\begin{table}
\centering
\scriptsize
\caption{Synthesis of research gaps from previous sections 4, 5, and 6.}
\label{tab:recommendations}
\begin{tblr}{
  width = \linewidth,
  colspec = {Q[35]Q[215]Q[692]},
  row{1} = {c},
  cell{2}{1} = {c},
  cell{2}{2} = {c},
  cell{3}{1} = {c},
  cell{3}{2} = {c},
  cell{4}{1} = {c},
  cell{4}{2} = {c},
  cell{5}{1} = {c},
  cell{5}{2} = {c},
  cell{6}{1} = {c},
  cell{6}{2} = {c},
  cell{7}{1} = {c},
  cell{7}{2} = {c},
  cell{8}{1} = {c},
  cell{8}{2} = {c},
  cell{9}{1} = {c},
  cell{9}{2} = {c},
  cell{10}{1} = {c},
  cell{10}{2} = {c},
  cell{11}{1} = {c},
  cell{11}{2} = {c},
  cell{12}{1} = {c},
  cell{12}{2} = {c},
  cell{13}{1} = {c},
  cell{13}{2} = {c},
  cell{14}{1} = {c},
  cell{14}{2} = {c},
  cell{15}{1} = {c},
  cell{15}{2} = {c},
  cell{16}{1} = {c},
  cell{16}{2} = {c},
  cell{17}{1} = {c},
  cell{17}{2} = {c},
  cell{18}{1} = {c},
  cell{18}{2} = {c},
  cell{19}{1} = {c},
  cell{19}{2} = {c},
  cell{20}{1} = {c},
  cell{20}{2} = {c},
  cell{21}{1} = {c},
  cell{21}{2} = {c},
  cell{22}{1} = {c},
  cell{22}{2} = {c},
  cell{23}{1} = {c},
  cell{23}{2} = {c},
  cell{24}{1} = {c},
  cell{24}{2} = {c},
  hline{1,25} = {-}{0.08em},
  hline{2} = {-}{},
}
\textbf{ID} & \textbf{Research Gap}                                     & \textbf{Description}                                                                                                                                                                                                  \\
1           & LLMs in Domains with Privacy Restrictions                 & To explore the use adoption of LLMs in domains that impose restrictions on privacy, such as government, finance, and healthcare. (See section \ref{rc:restrictDomains})                                   \\
2           & Studies in Industry Settings                              & Most of the studies were conducted in university settings. There is a need for case studies in industry settings. (See \ref{rc:industrySettings})                                                    \\
3           & LLMs supporting Effort Estimation and Task Prioritisation & To explore the potential for LLMs support novice developers in~effort estimation and task prioritisation.~~(See section \ref{rc:effortEstimationAndTaskPrioritisation})                              \\
4           & Open-source LLMs in SE Education                          & To investigate the potential benefits of novice developers using open-source LLMs in SE education. (See section \ref{rc:opensourceLLMsAndSEEducation})                                               \\
5           & LLMs and Domain Expertise                                 & To study the potential influence of LLMs on junior developers quickly improving their domain expertise, such as finance and retail.~ (See section \ref{rc:domainExpertise})                            \\
6           & LLMs and Architecture Definition~                         & To explore the potential of LLMs to improve novice developers' understanding of architecture definition. (See section \ref{rc:architectureDefinition})                                               \\
7           & Content creators and AI companies                         & To investigate the perspective of tech content creators, focusing on achieving a solution that balances the interests of content creators and AI companies. (See section \ref{rc:balacingInterests}) \\
8           & LLMs and Novices' Companies' Onboarding                   & To study the impact of junior developers using LLMs during companies' onboarding. (See section \ref{rc:companiesOnboard})                                                                            \\
9           & LLMs influencing Monocultures~                            & To explore the potential influence of LLMs on the generation of monocultures in novice developers. (See section~\ref{rc:monoculture})                                                                \\
10          & LLMs on Mentorship Interactions                           & To explore the consequences of LLM adoption on mentorship interactions between senior and novice developers. (See section \ref{rc:mentorship})                                                            \\
11          & LLMs and Developers' Code Reading Skills                  & To investigate the consequences on novice developers' code reading skills by adopting LLMs. (See section \ref{rc:readingSkills})                                                                     \\
12          & LLMs and Job Interviews                                   & To study the impact of LLM users in tech job interviews. (See section \ref{rc:techJobInterview})                                                                                                     \\
13          & LLMs and~                                                 & To explore how LLM users would experience migrating to other programming languages.~(See section \ref{rc:learnOtherProgLang})                                                                        \\
14          & Unique and Emerging LLM Uses Cases                        & To investigate interesting use cases of using LLMs by novice developers, such as for simulating user interaction. (See section \ref{rc:uniqueUseCase})                                               \\
15          & LLMs for Code Review                                      & To study the use of LLMs by novice developers for code review. (See section \ref{rc:codeReview})                                                                                                     \\
16          & LLMs in Under-explored Topics                             & To investigate the potential consequences of LLM adoption in underexplored topics, such as the impact on work culture. (See section \ref{rc:underploredTopics})                                    \\
17          & Co-variables on LLM adoption                              & To explore the potential co-variables on LLM adoption, such as work culture and satisfaction. (See section \ref{rc:coVariables})                                                                      \\
18          & Outsourcing to LLMs                                       & To study what potential software development tasks novice developers would be comfortable outsourcing to LLMs. (See section \ref{rc:outsourcing})                                                    \\
19          & LLMs Instigating Self-Learning                            & To investigate the potential of promoting self-learning in junior software developers (See section \ref{rc:selfLearning})                                                                            \\
20          & LLM and Impostor Phenomenon~                              & To explore the influence on LLMs to support novice developers overcome impostor phenomenon (See section~\ref{rc:impostor})                                                                           \\
21          & LLMs in Software Quality Education                        & To study how the adoption of LLM in education can improve CS/SE students' code quality. (See section~\ref{rc:codeQuality})                                                                           \\
22          & Experiment with LLMs                                      & To investigate the applicability of adjusting LLMs to act proactively in correcting novice developers' mistakes. (See section \ref{rc:adjustingLLMsToActActively})                                   \\
23            & Software Team Leaders' Expectations                       & To explore software team leaders' and IT managers' expectations for novice developers in an LLMs era. (See section \ref{rc:expectations})                                                            
\end{tblr}
\end{table}

\textbf{Ethnography Studies with Early Career Developers.} During our analysis of the research methods employed in the primary studies (See Fig. \ref{fig:researchMethodOverType}), we observe that there is no paper using ethnography as a research method. According to Sharp et al. \cite{sharp:2016}, SE researchers can leverage ethnography to not only investigate what software practitioners do but also their motivation behind it. However, although this research method is widely employed in Computer-Supported Cooperative Work \cite{blomberg:2013} and Human-Computer Interaction \cite{blomberg:2009}, it is still not widely adopted by SE researchers, as traditional research methods (e.g., interviews and questionnaires) are \cite{sharp:2016}. In the context of LLMs, de Seta et al.\cite{de:2024} proposed \textit{synthetic ethnography} as an adaptation of the traditional ethnography focused on qualitative studies of LLMs. We suggest that future research follow this ethnography variation, as exemplified in the study by Holmquist et al. \cite{holmquist:2025}.

\textbf{LLMs Customised to Novice Developers' Needs.} We presented in Section \ref{sec:rq3} many positive and negative aspects associated with novice developers using LLM tools. However, we also observed that current LLM tools do not provide mechanisms to effectively lead novice developers in improving their skills, as they can simply copy and paste AI suggestions. For this reason, we suggest that future studies investigate approaches that combine educational potential with principles of productivity and efficiency, which are essential in industry. Otherwise, if these mechanisms are not developed and novice developers become overly reliant on LLM tools, the improvement of their skills (e.g., critical thinking \cite{lee:2025}, creativity \cite{kumar:2025}) could be hindered.

\section{Threats to Validity} \label{sec:threatsToValidity}

Although this SLR follows the guidelines presented by Kitchenham and Charters \cite{kitchenham:2007, kitchenham:2022}, and the methodology is the result of many discussions involving experienced researchers, it is essential to acknowledge potential threats to the validity of the results (e.g., missing relevant papers). Ampatzoglou et al. \cite{ampatzoglou:2019} highlight that the empirical SE community adopts Wohlin et al.'s approach \cite{wohlin:2012} (\textit{conclusion}, \textit{internal}, \textit{construct}, and \textit{external validity}) for quantitative research within software engineering.

\textbf{Internal Validity.} It examines whether a study condition makes a difference or not, and whether there is sufficient evidence in the data collected that supports the study conclusions \cite{ampatzoglou:2019, wohlin:2012}. This study methodology was initially developed by the first author, but it was also reviewed by the other co-authors and SLR experts before starting the study. Then, the first author conducted a pilot test for the paper selection, which enabled the identification of the need for adjusting the search string and inclusion and exclusion criteria after validating with the other authors. The search string was customised for each one of the 7 data sources, being built on several attempts targeting to optimise the results. Our SLR protocol encompasses various rounds of filtering the studies,  focusing on minimising the selection bias. A pilot test was also conducted for the data extraction process, which enabled us to adjust after identifying other relevant data that could be extracted from the studies.

\textbf{Construct Validity.} It refers to how well constructed the methodology is, being effective to what was intended \cite{ampatzoglou:2019, wohlin:2012}. Our SLR protocol combines different strategies to ensure that we are collecting relevant studies: a) primary search: involving the six most relevant digital libraries; b) secondary search (\textit{forward} and \textit{backward snowballing} as secondary search); c) search in the grey literature (i.e., arXiv). The pilot test for paper selection and data extraction also helped to evaluate and improve the SLR protocol (e.g., inclusion and exclusion criteria), making the necessary adjustments. Finally, continuous discussions between the authors (and other SLR experts) helped to build a well-defined SLR protocol while reducing threats.  

\textbf{Conclusion Validity.} This aspect refers to which extension the conclusions can be reached from the data collected \cite{ampatzoglou:2019, wohlin:2012}. Focusing on a solid data extraction process, our data extraction form was developed based on our RQs, which were developed using the PICOC framework. The SLR data, including the spreadsheets with the stages accomplished throughout this SLR, allow other researchers to replicate this study.

\textbf{External Validity.} It examines whether the findings can be generalised \cite{ampatzoglou:2019, wohlin:2012}. Our search was carefully adjusted to include publications from 2022, aligning with ChatGPT's public release. However, we did not limit our selection to studies using ChatGPT specifically. This SLR also follows Naveed et al. \cite{naveed:2024} while restricting our search to publications in the English language because it is broadly adopted for reporting research studies. We also excluded vision and opinion papers since they may highlight an individual perspective and lack generalisation. Finally, our findings were based on 80 studies - a reasonable amount considering the empirical software engineering literature (e.g., \cite{naveed:2024, khalajzadeh:2024, hidellaarachchi:2021}).

\section{Conclusion} \label{sec:conclusion}

We conducted a systematic literature review on novice software developers' perspectives on the adoption of LLM tools, which resulted in the selection of 80 relevant studies by following the guidelines provided by Kitchenham et al. \cite{kitchenham:2007, kitchenham:2022}. 

 Our analysis of our primary studies indicates that: 1) there is a fast growth in publications in LLM4SE; 2) Although initially, research focused basically on ChatGPT and Copilot, more diversity in LLMs is emerging in the recent publications; 3) there are few papers focused on novice developers in industrial settings; 4) Participants in most of the selected studies have a mix of positive and negative perceptions about the impact of adopting LLM tools.

In RQ1, we identified the motivations and methodological approaches of the studies. Interviews and questionnaires are the research methods more commonly adopted in the primary studies. In RQ2, we identified the software development tasks and LLMs used by novice developers in the studies. While most of the primary studies mention SE tasks related to software development and software quality assurance, there is a research gap in studies that explore software management-related tasks. In RQ3, we identified a variety of topics discussed in the studies (e.g., emotions, output quality) as well as benefits, challenges, and recommendations. Not surprisingly, gains in productivity and efficiency are the most commonly reported benefits. We found many challenges potently indicating that novices are not ready for LLMs, and many recommendations suggesting that developers be cautious when adopting these tools. Finally, we present the study limitations and future research needs in RQ4. Most of the studies have limitations regarding data collection and analysis, and exploratory studies are suggested for future investigations. For future research, we intend to investigate the research gaps presented in section \ref{sec:discussion} (See Table \ref{tab:recommendations}).

%%
%% The acknowledgments section is defined using the "acks" environment
%% (and NOT an unnumbered section). This ensures the proper
%% identification of the section in the article metadata, and the
%% consistent spelling of the heading.
% \begin{acks}
% To Robert, for the bagels and explaining CMYK and color spaces.
% \end{acks}

\bibliographystyleP{unsrt}
\bibliographyP{primary-studies}

%%
%% The next two lines define the bibliography style to be used, and
%% the bibliography file.
\bibliographystyle{ACM-Reference-Format}
\bibliography{base}
% \bibliography{base}

%%
%% If your work has an appendix, this is the place to put it.

\appendix

% \subsection{Part One}

\end{document}